# A signal discovery step in interstellar communication

*William J. Crilly Jr.*

Green Bank Observatory, West Virginia, USA

***Abstract*—Prior work using synchronized, geographically spaced radio telescopes, and a radio interferometer, suggests that narrow-bandwidth polarized pulse pair measurements repeatedly falsify a noise-cause hypothesis, given a prior celestial direction of interest. A four-step method was proposed, tested, and reported, using interferometer phase measurements, to seek common celestial directions among pulse pair components, during 92 days of observation. In the work reported here, the proposed four-step signal discovery method is simplified to have a single step. A 123.8 day interferometer experiment provides measurement evidence supporting a hypothesis that the prior direction of interest, and a second direction of interest, are associated with celestial coordinates. Each pointing direction measures statistical power at greater than six standard deviations, with some indications of associated interferometer-induced Right Ascension aliasing. Explanations are proposed and discussed.**

## I. Introduction

**Prior work**

Polarized narrow-bandwidth pulse pairs, possibly of celestial origin, at Right Ascension (RA) 5.25 ± 0.15 hr and Declination (DEC) -7.6º±1º, have been repeatedly observed to have high statistical power of increased pulse pair count, using synchronized, geographically-spaced radio telescopes, **[1]** a single radio telescope, **[2][3][4]** and a radio interferometer **[5][6][7]**. In a previous report, **[7]** Fig.1, a second anomalous pulse pair count was observed near RA 8.9 hr and DEC -4.3º±1º, with Full Width Half Maximum (FWHM) interferometer element at 5.3º±1º.

**Objectives**

The objectives of the current work are to present long duration interferometer observations of hypothetical celestial narrow-bandwidth polarized pulse pairs, simplify signal discovery methods, and improve the testability of a celestial origin hypothesis.

## II. Hypothesis

**Hypothesis**

Celestial polarized narrow-bandwidth pulse pairs, observed using a radio interferometer, are expected to indicate high statistical power during long duration pulse pair counting experiments. Pulse pairs having a celestial origin are expected to indicate anomalous pulse pair counts in directions of prior and aliased interferometer response. Celestial pulse pair production mechanisms are hypothesized to have non-Additive White Gaussian Noise (AWGN) properties.

**Falsification**

The celestial-cause hypothesis is conditionally falsified if primary and aliased responses indicate statistical power consistent with the statistical power of a comparison group. Natural object and RFI hypotheses are conditionally falsified if reasonable models cannot explain pulse pair count and associated measurements.

**Comparison group**

RA directions offset from directions of interest are used to provide a comparison group that is expected to indicate lower statistical power effect size, and that may be explained by AWGN.

**Theoretical pulse pair production mechanisms**

Polarized narrow-bandwidth pulse pairs may be observed due to many possible causes, including Radio Frequency Interference (RFI), equipment issues, natural astronomical objects, noise and extraterrestrial civilization transmitters. Conclusions about transmission mechanisms may be proposed and considered after each mechanism is modeled, studied, experimentally tested and replicated.

## III. Method of Measurement

The primary measurement used in this work is the difference in measured RF phase $\phi$ across interferometer antenna elements, of two $\Delta$f radio frequency-spaced, 3.7 Hz bandwidth, 0.27 s duration, simultaneous pulses in a hypothetical transmitted pulse pair. The phase is compensated for antenna element to phase detector measurement delay difference using

$$\Delta_{\Delta f}\,\Delta_{EW}\phi = \Delta_{\Delta f}\,\Delta_{EW}\phi_{MEASURED} + 2\pi\,\Delta f\,\tau_{INT}\,, \qquad (1)$$

where the $\Delta_{\Delta f}$ and $\Delta_{EW}$ calculations produce phase difference across pulse pair difference frequency, and antenna elements, respectively, $\Delta$f is the frequency difference between simultaneous pulses in candidate pulse pairs, $\tau_{INT}$ is the measured difference in delay between signal paths from the two antenna elements to a complex cross-correlator and 3.7 Hz bandwidth 0.27 s averaging measuring receiver.

**Measurement changes in the current work**

To reduce the possibility of instrument adjustment bias, experimental settings in the reported O7a and O7b observation runs **[7]** have been retained in the current report of O7a, O7b and O8a observation runs, with exceptions to set the frequency range limits to 1398 and 1451 MHz, and RA bin size decreased from 0.015 hr RA to 0.0075 hr.

RA and MJD coverage has been expanded in the current report. A new O8a dataset comprises first-level processed interferometer data files that were not second-level processed in the previous O7a and O7b report **[7]**. First-level processed files were added to flatten the event probability curve across an RA range wider than that in the previous report, and to add MJD days subsequent to the O7a and O7b report.

The second-level processing discovery method in the current work implements only the first step in the method described in the O7a and O7b report **[7]**.

**Measurement system**

Two ten foot diameter offset-fed parabolic interferometer antennas, 3.7 Hz bandwidth receivers, FX cross-correlator **[8]**, pulse pair first-level storage of 3.7 Hz bandwidth, 0.27 s duration pulses exceeding 8.5 dB signal-to-noise ratio (SNR), second-level processing and RFI amelioration are described in the O7a and O7b report and previous reports **[5][6][7]**. The antennas are located in a rural forested area having controlled RFI. The second-level processing machine implements communication signal detection algorithms and presentation of results in graphics files. The DEC was set to -4.3º±1º throughout the experiment. Polarization is right hand circular.

William J. (Skip) Crilly Jr. is a Volunteer Science Ambassador in Education & Public Outreach of the Green Bank Observatory. email: wcrilly@nrao.edu

**Theory of operation**

Polarized pulses exceeding an SNR threshold are RA binned to a resolution of 0.0075 hr RA, limiting the expected $\Delta_{EW}\phi$ RA-induced range to $\pm\pi$ radians divided by 15.6, the number of RA bins per RA aliasing period. RF frequency and instrument East-West differential delay are not used to correct the $\Delta_{EW}\phi$ measurement. Therefore, measured values of $\Delta_{EW}\phi$ are expected to have a $\pm\pi$ radians sawtooth response over RF frequency, based on the -82 ns differential East-West instrument delay, having a 12.2 MHz RF frequency period. In combination, RA binning, RF frequency and phase noise produce a subset of near-zero $\Delta_{EW}\phi$ values. A $\Delta_{EW}\phi$ filter is set in this experiment at 0.0 ±0.1 radians.

A pair of simultaneously transmitted narrow bandwidth pulses, received within an RA bin, i.e. pulses in a pair spaced in frequency by $\Delta f$, are expected to measure the same $\Delta_{EW}\phi$, albeit affected by $\Delta f$ and phase noise sources. The measured difference in phase between simultaneous pulses in the pair, $\Delta_{\Delta f}\Delta_{EW}\phi_{MEASURED}$, is corrected using the measured $\tau_{INT}$ and pulse pair $\Delta f$ value, using Equation (1), to produce a $\Delta_{\Delta f}\Delta_{EW}\phi$ value expected to be centered near zero radians, while having three phase noise sources, quantized RA bin-induced phase noise at $\pm\pi/15.6 \approx \pm0.2$ radians, SNR=10 phase noise at $\pm\tan^{-1}(10^{-0.5}) \approx \pm0.3$ radians, and pulse pair $\Delta f$=7 MHz residual ±1 ns $\tau_{INT}$-uncertainty phase noise ≈ ±0.1 radians. Combined, the uncorrelated root mean square phase noise calculates to ±0.37 radians. The $\Delta_{\Delta f}\Delta_{EW}\phi$ filter limit in second-level processing is set to 0.0 ± 0.8 radians.

Sorting the candidate pulse pair data set by increasing value of $|\Delta_{\Delta f}\Delta_{EW}\phi|$ before applying a binomial model to calculate the AWGN-expected mean count of pulse pairs in an RA bin, provides a means to seek anomalous counts of pulse pairs in RA bins. The results are presented as Cohen's **d** effect size **[9],** the pulse pair count mean shift from the binomial AWGN-cause mean, measured in binomial-AWGN model standard deviations.

## IV. Observations

The O7a and O7b observation dataset, covering 92 days of first-level processed data, were second-level processed for a previous report.**[7]** O7a and O7b datasets were combined with O8a, i.e. post-report days and increased RA coverage of pre-report days. Combined first-level processed files span 123.8 days, and have an average reported RA coverage of 4.8 hours per day. Given the antennas' FWHM measured beam-width at 5.3 degrees, the experiment reported here examines a celestial sky coverage of approximately 360 square degrees, repeatedly, during the 123.8 days.

The interferometer element baseline angle was measured at 180.0º±0.3º azimuth using Sun shadow measurements. Baseline physical length measured 33.0±0.1wavelengths at 1425 MHz, excluding potential feed illumination center offsets.

**Presentation of results**

**Figure 1** shows the Cohen's **d** effect size of pulse pair counts across RA bins, with MJD 60586 correlator and continuum measurements. **Figs. 2-3** show a higher RA resolution of the prior O1-O6 and O7a and O7b pulse pair directions of interest. **Figs. 4-7** show the 954 Hz and 50 MHz bandwidth noise power measurements, integrated over 0.27 s, at the time of each pulse pair event. **Figs. 8-10** show the Δf, RF frequency and MJD of each pulse pair event. **Fig. 11** shows pulse pair counts. **Figs. 12-13 s**how the $\log_{10}$ likelihood of the measured SNR of pulses and pulse pairs. **Figs. 14-17** show the $\Delta_{\Delta f}\Delta_{EW}\phi$ and $\Delta_{EW}\phi$ of pulse pairs and pulses. **Fig. 18 s**hows the binomial model event probability per RA bin. **Fig. 19** shows the MJD days observed during the experiment. **Fig. 20** extends the RA range of **Fig. 1**. **Fig. 21** shows modification of the detection algorithm to test for an expected null result, in a receiver system test.

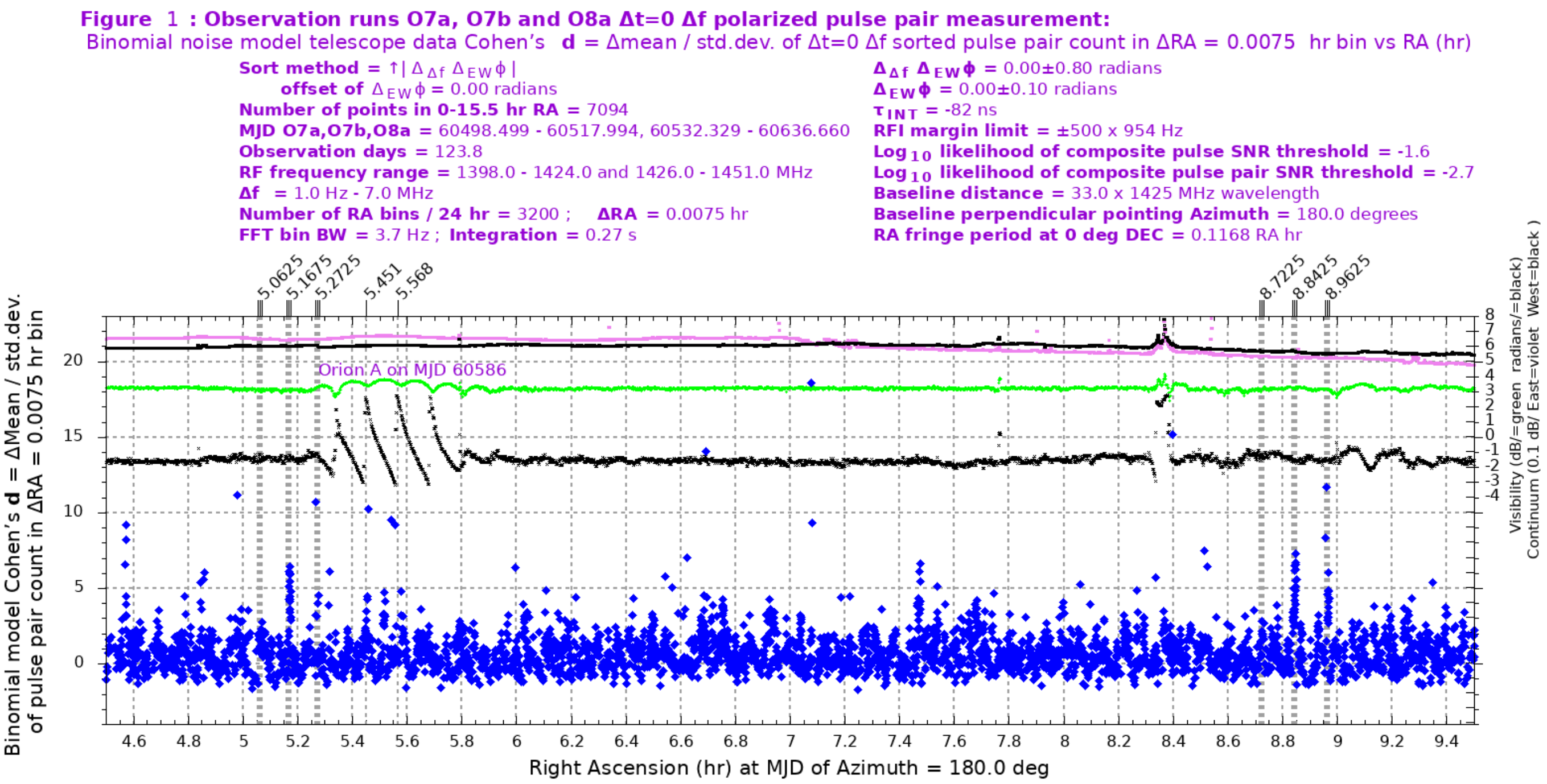

**Figure 1:** Continuum noise power and FX complex correlator measurements, observed on MJD 60586, are presented together with the statistical power of 3.7 Hz bandwidth, 0.27 s duration polarized pulse pairs, recorded during the 123.8 day experiment. The high correlator response at 5.5 hr RA is due to the Orion A emission nebula at -5.4º DEC. **[10][11]**. Instrument delay difference of -82 ns between interferometer elements was compensated by choosing to display the -5 x 16 ns correlator tap. The bandwidth of the continuum and correlator measurements is approximately 50 MHz, at 0.27 s integration. The time-dependent geometric space delay was not compensated, resulting in the approximate sawtooth phase response across RA, indicating a period of 0.117 hr RA. The prior pointing direction of interest indicates greater than 6 standard deviations Cohen's **d** in the 5.1675 ± 0.00375 hr RA bin, within the prior range, 5.25 ± 0.15 hr RA. A second apparent response is indicated in two adjacent bins centered at 8.8425 hr RA, with apparent RA aliasing at ±0.12 hr RA. **Figs. 2** and **3** show detail.

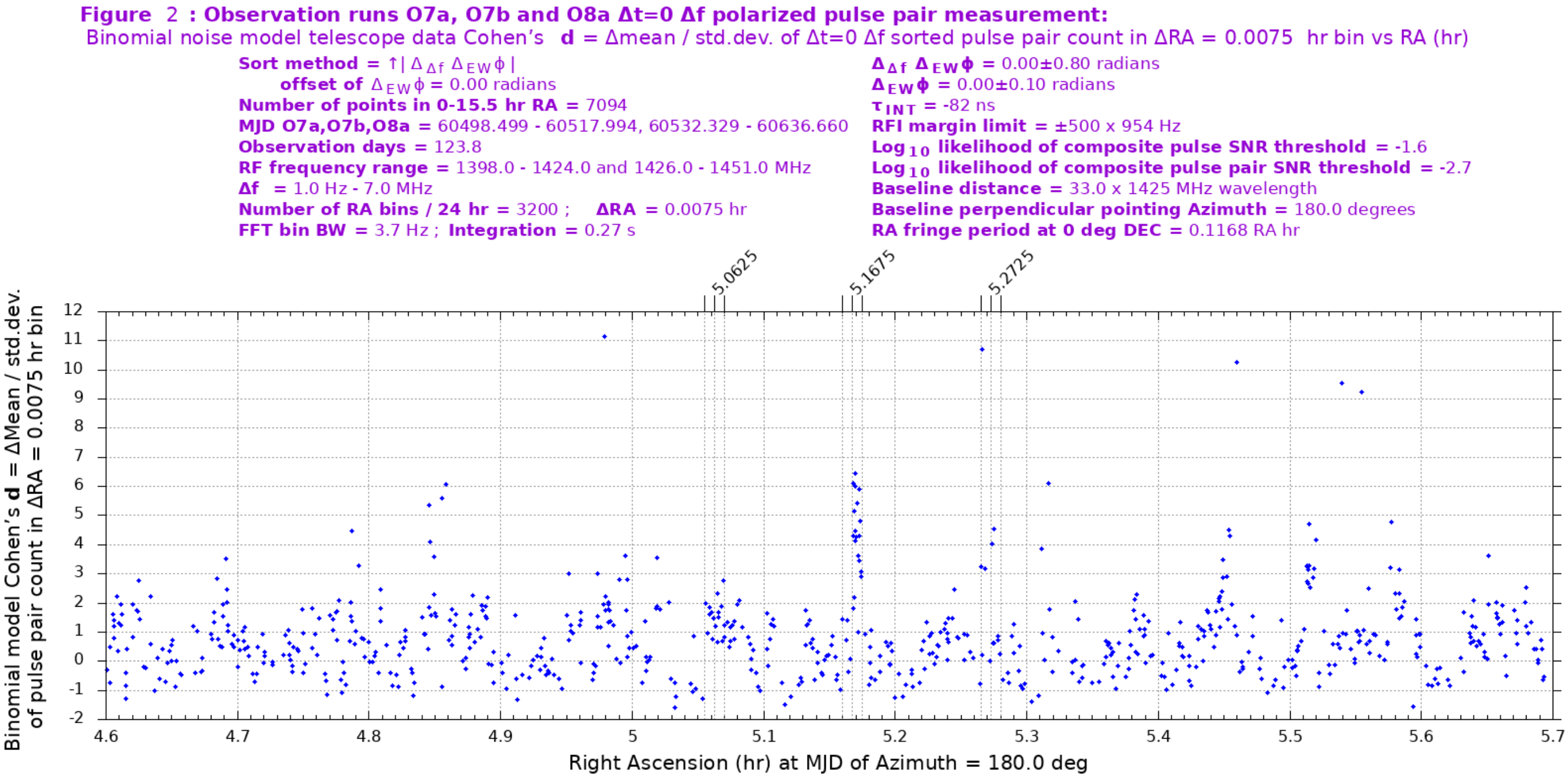

**Figure 2:** Cohen's **d** measures the effect size, across differing populations, of a mean shift in a modeled process. **[9]** Nineteen prior direction pulse pairs appear in RA bin 5.1675-5.175 hr, between 1 and 7 standard deviations, implying the presence of pulse pairs not explained by an AWGN cause. Pulse pairs at an offset of 0.10875 hr RA below the large group of pulse pairs at 5.17125 hr. are observed, and might be due to interferometer RA-aliasing of one wavelength of interferometer geometric space delay. Five pulse pairs appear near 5.2725 hr RA, at >4 standard deviations, one at 10.7 standard deviations. Orion A fringe period measures 0.117 hr RA aliasing, and physical antenna geometry measures 0.1168 hr RA at 0º DEC. The aliased RA bins are plotted at a spacing of ±0.105 hr RA from the central point of the two bins at 5.1675 hr RA. The difference between the calculated 0.1168 hr RA aliasing value and suspected RA aliasing may be due to measurement uncertainty, phase noise **[7],** low statistical power of the -0.10875 hr RA group, and sparse data in the suspected RA alias groups.

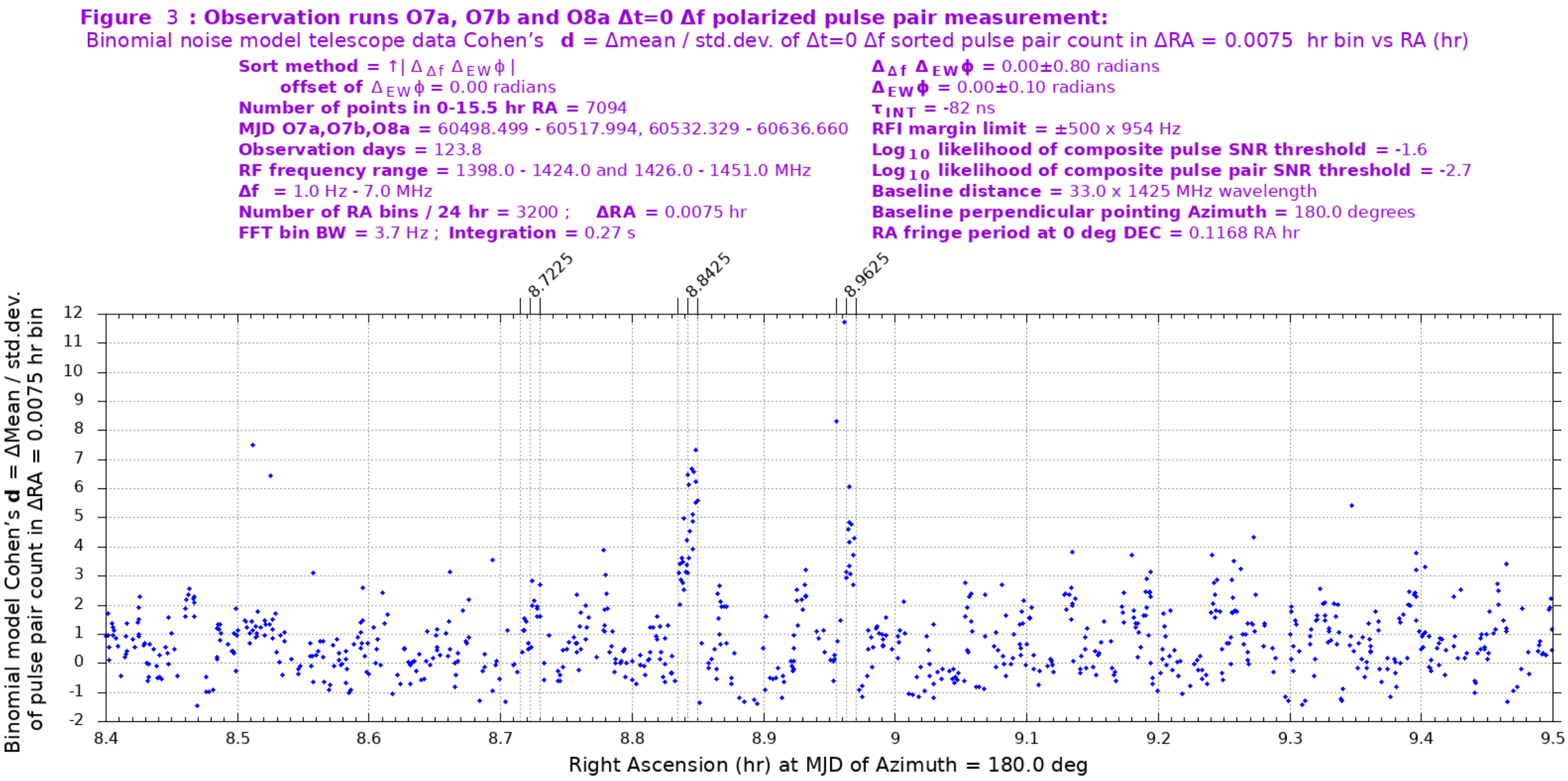

**Figure 3:** Twenty-seven pulse pairs are observed in two adjacent RA bins, centered at 8.8425 hr RA, measuring between 2.0 and 7.3 standard deviations of Cohen's **d**. Pulse pairs in ±0.120 hr RA adjacent bins, implies the potential presence of a 3.7 Hz bandwidth celestial source transmitting Δf RF frequency-spaced polarized pulse pairs, during the 123.8 day experiment. An RA bin corresponds to a measurement time window of 27 seconds per day when pointing along the celestial equator. Interferometer elements have beam center pointing at -4.3º±1º DEC with a 5.3º±1º FWHM. Known RFI does not seem to easily produce the precise RA measurements observed. **Figs. 4-7** and **Figs. 20-21** investigate ideas that the 3.7 Hz bandwidth pulse pairs might be high SNR outlier components of 954 Hz and/or 50 MHz occupied bandwidth emissions due to equipment issues, RFI or celestial objects.

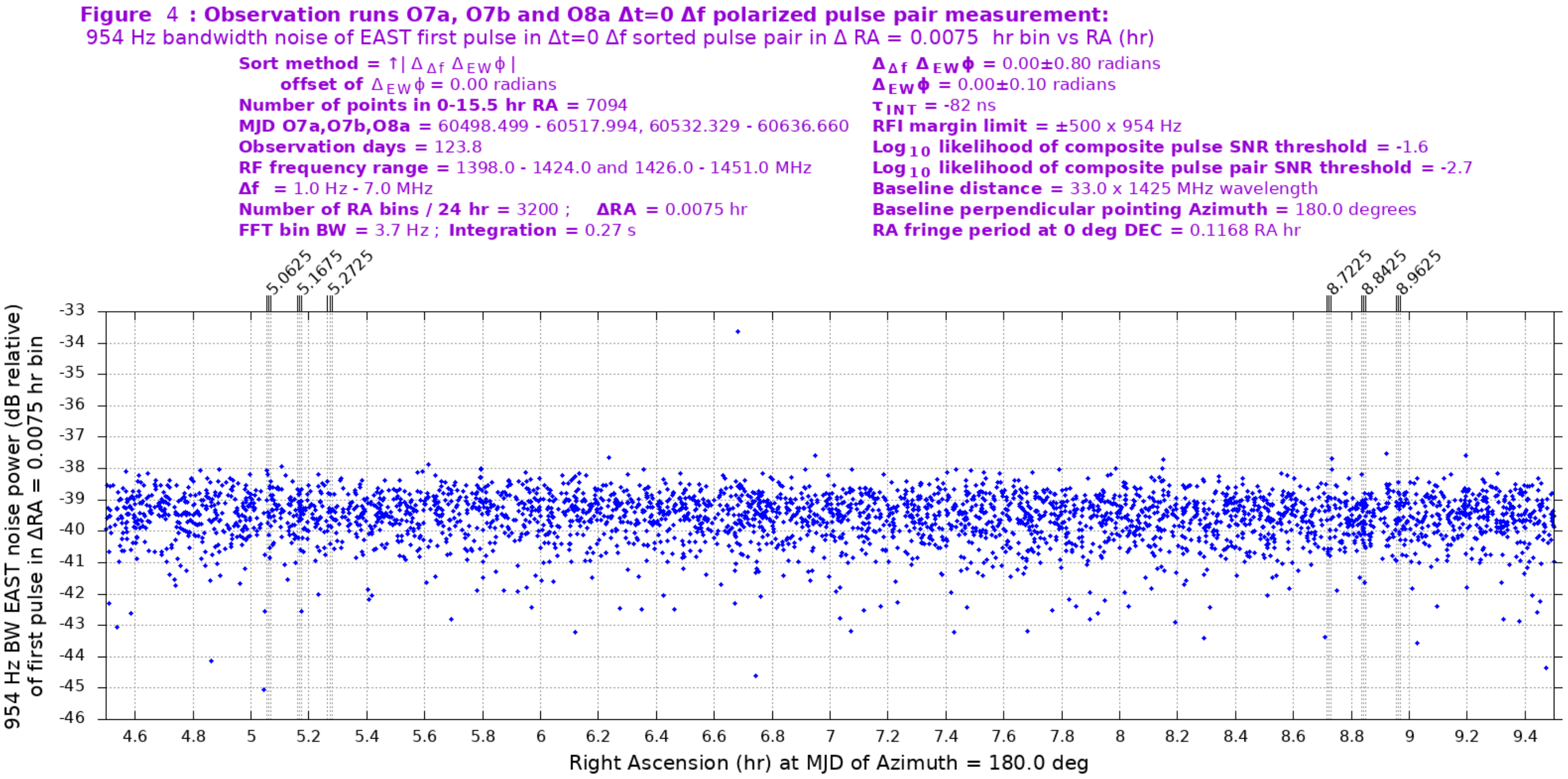

**Figure 4:** The 954 Hz bandwidth noise power of each interferometer element antenna signal is measured at the time and RF frequency of each recorded pulse SNR threshold-crossing event, in first-level processing. This measurement is useful in determining if the pulse pair event may be caused by equipment, celestial object emission, or RFI. Evidence does not appear in East element data that implies that the pulse events occur at the time of increased power 954 Hz bandwidth emissions. Rather, the 3.7 Hz bandwidth pulse pairs observed in the directions of interest appear to be isolated narrow bandwidth signals.

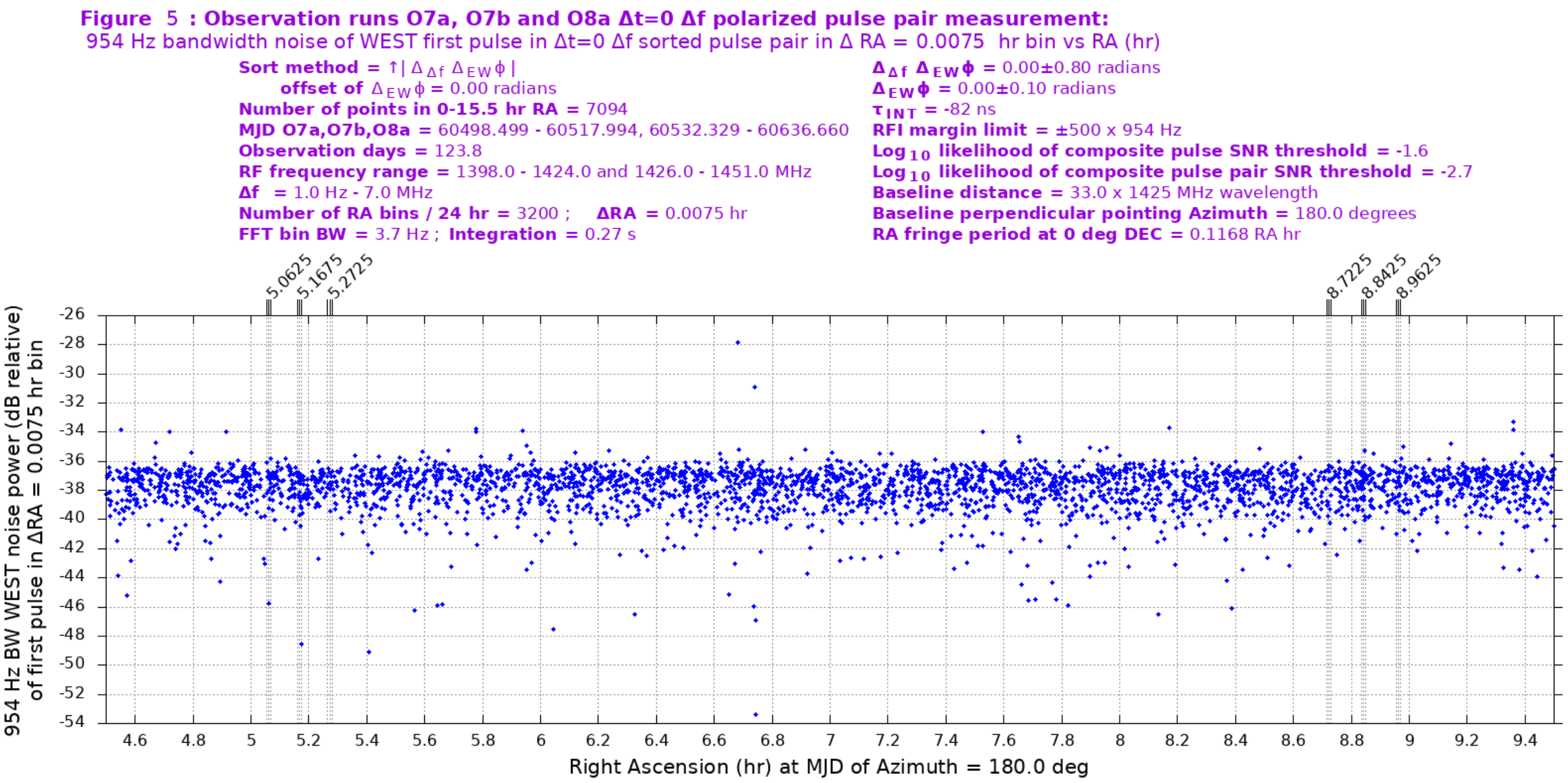

**Figure 5:** The West antenna element does not indicate significant 954 Hz bandwidth noise power measurements, e.g. greater than 3 dB above average, repeatedly in RA bins, at the times and RF frequency of the anomalous 3.7 Hz bandwidth pulse pair events. Isolated narrow bandwidth pulses implies a possibility that the observed pulse pairs may be intentional increased-power outlier components of a wide bandwidth communication signal. **[1]** Such signals are expected in communications systems designed to be easily discovered. 50 MHz bandwidth power measurements are presented in **Figs. 6 and 7**.

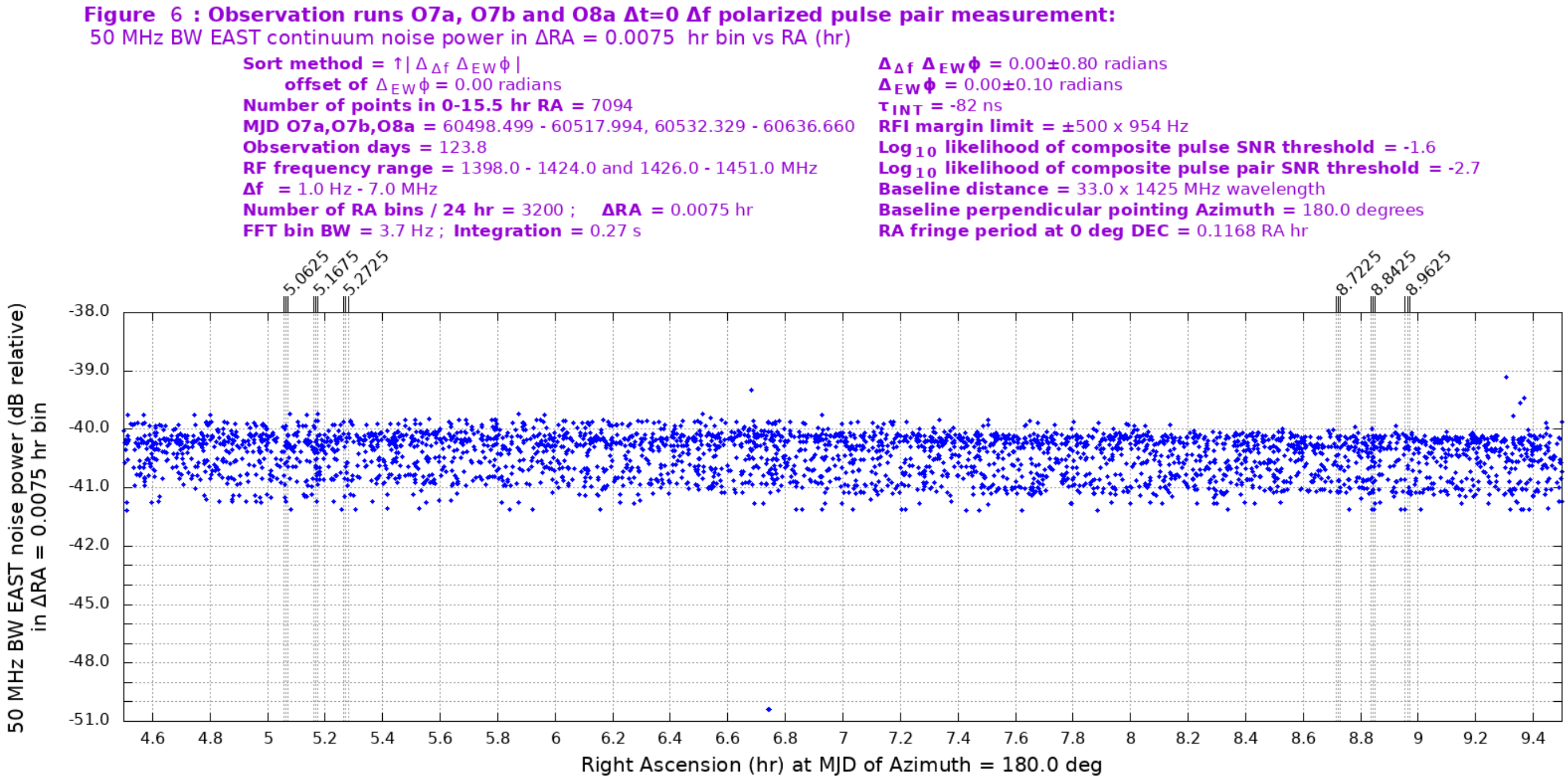

**Figure 6:** Wide bandwidth measurements are important because they possibly indicate the presence of local RFI, equipment problems and bursts of celestial energy. The measurement system makes simultaneous wide bandwidth measurements of each antenna element, as pulses are recorded in first-level signal processing. Due to the increased counts of pulse pairs in the directions of interest, some outliers showing higher power measurements may be expected. The short-term 50 MHz power measurement variation is typically a few thousandths of a dB, due to high bandwidth-time product. The variation shown here is much higher, due to continuum measurement variation over MJD days, sky noise and temperature. Study of wide bandwidth power measurements, associated with pulse pair events, is a topic of further work.

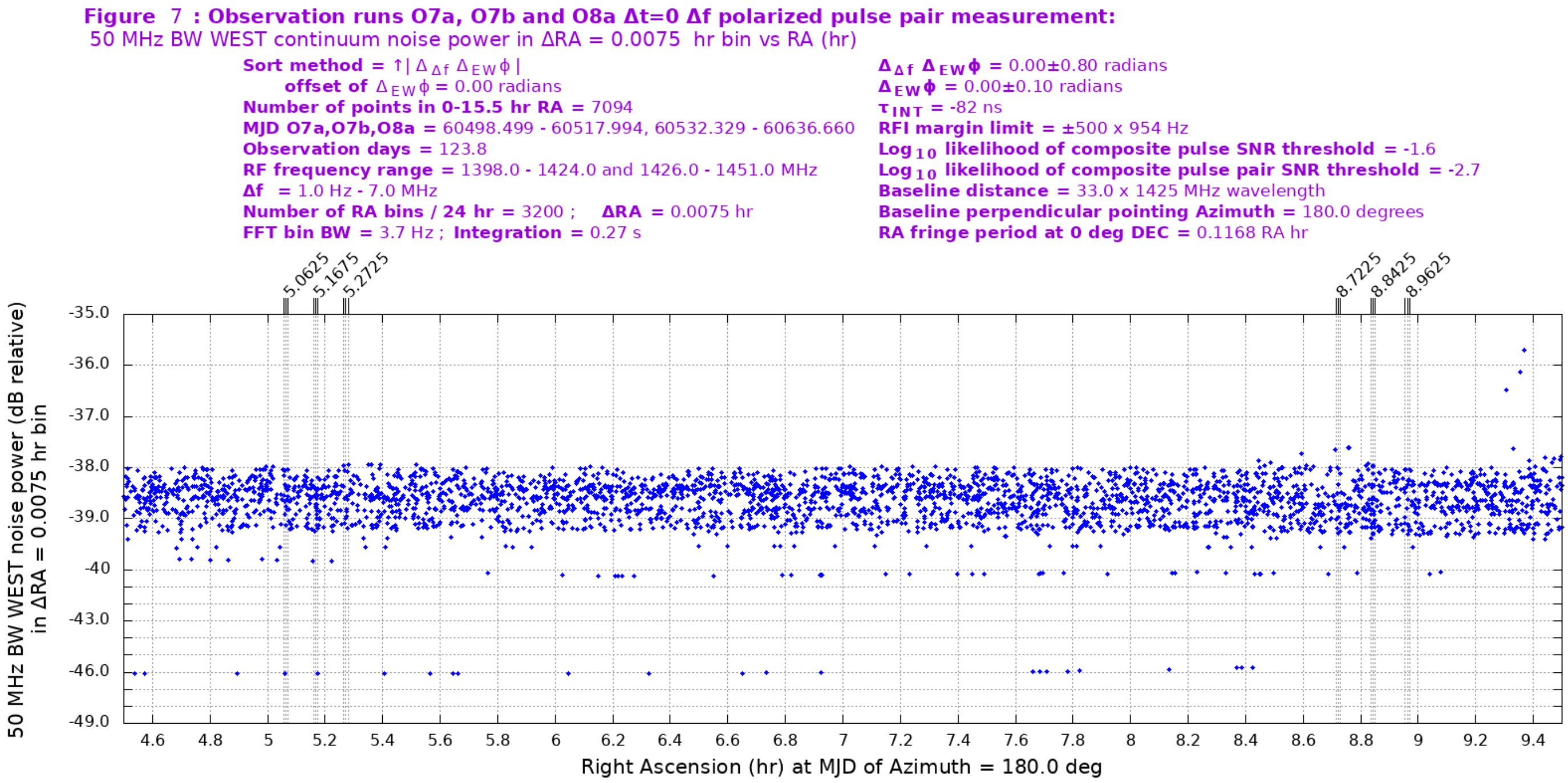

**Figure 7:** The absence of synchronous 954 Hz and 50 MHz bandwidth anomalous high noise power measurements suggests that the observed 3.7 Hz bandwidth pulse pairs are not random outliers of signals having these bandwidths. In AWGN, Rayleigh amplitude, and exponential power distributed density functions are expected. Some wide bandwidth emissions, e.g. RFI, have the potential to contain narrow bandwidth high SNR components. Measurements of the comparison group, i.e. RA bins not in the directions of interest, are thought to be measurements of filtered system and sky noise, and may be used to estimate the AWGN-expected narrow bandwidth pulse pair count. Further work is required to seek pulsed power in other bandwidths, between 3.7 Hz and 50 MHz, at the time of the hypothetically received 3.7 Hz bandwidth pulse pairs. Higher sensitivity radio telescopes are expected to help provide a better understanding of a potential association of wide bandwidth and narrow bandwidth emissions.

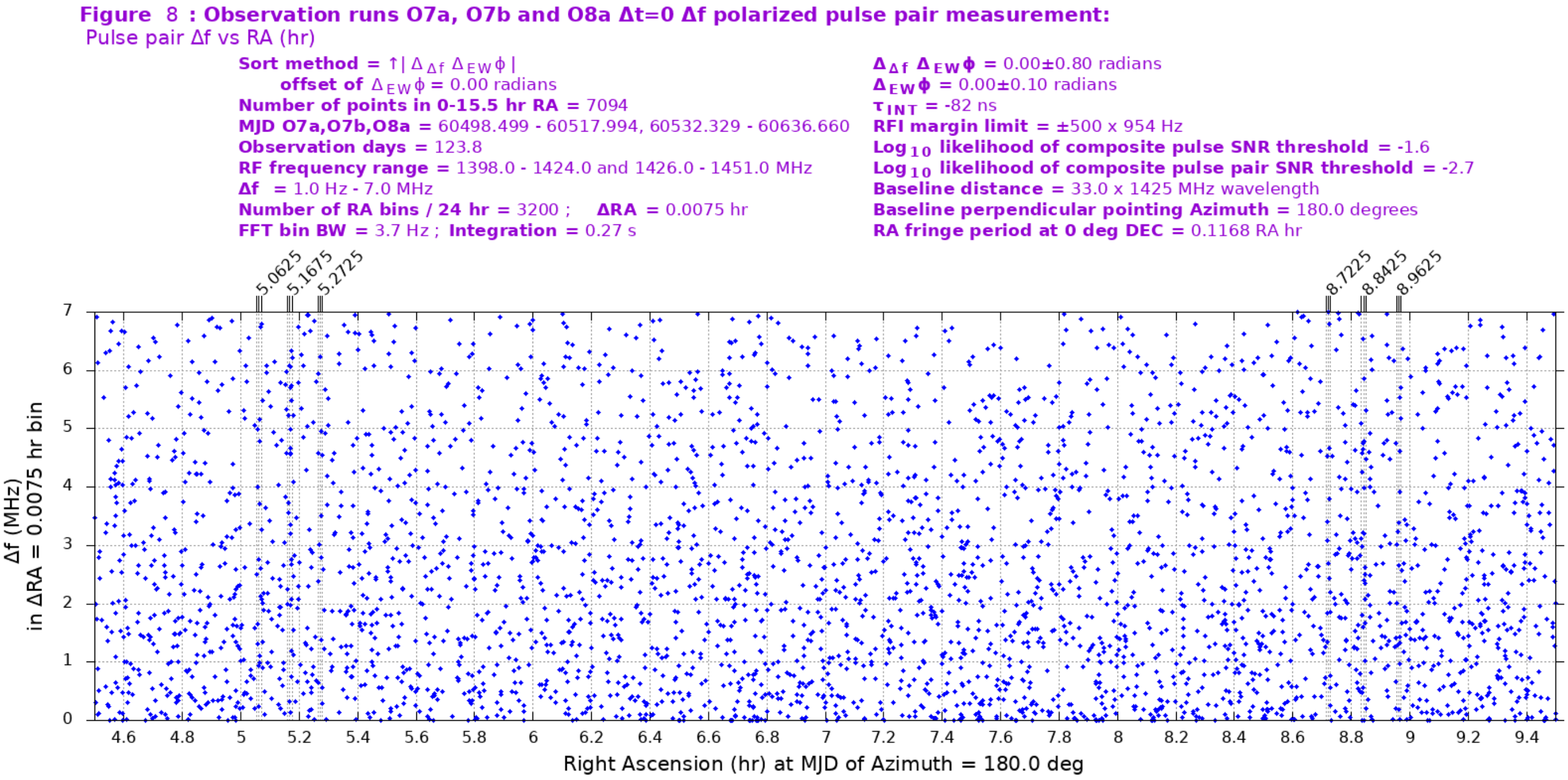

**Figure 8:** The measurement of the frequency spacing of simultaneous 3.7 Hz, 0.27 s duration pulses within a pulse pair, Δf, helps one investigate if an RFI or equipment issue might be causing the pulse pairs. No apparent concentration and/or quantization of Δf values is observed.

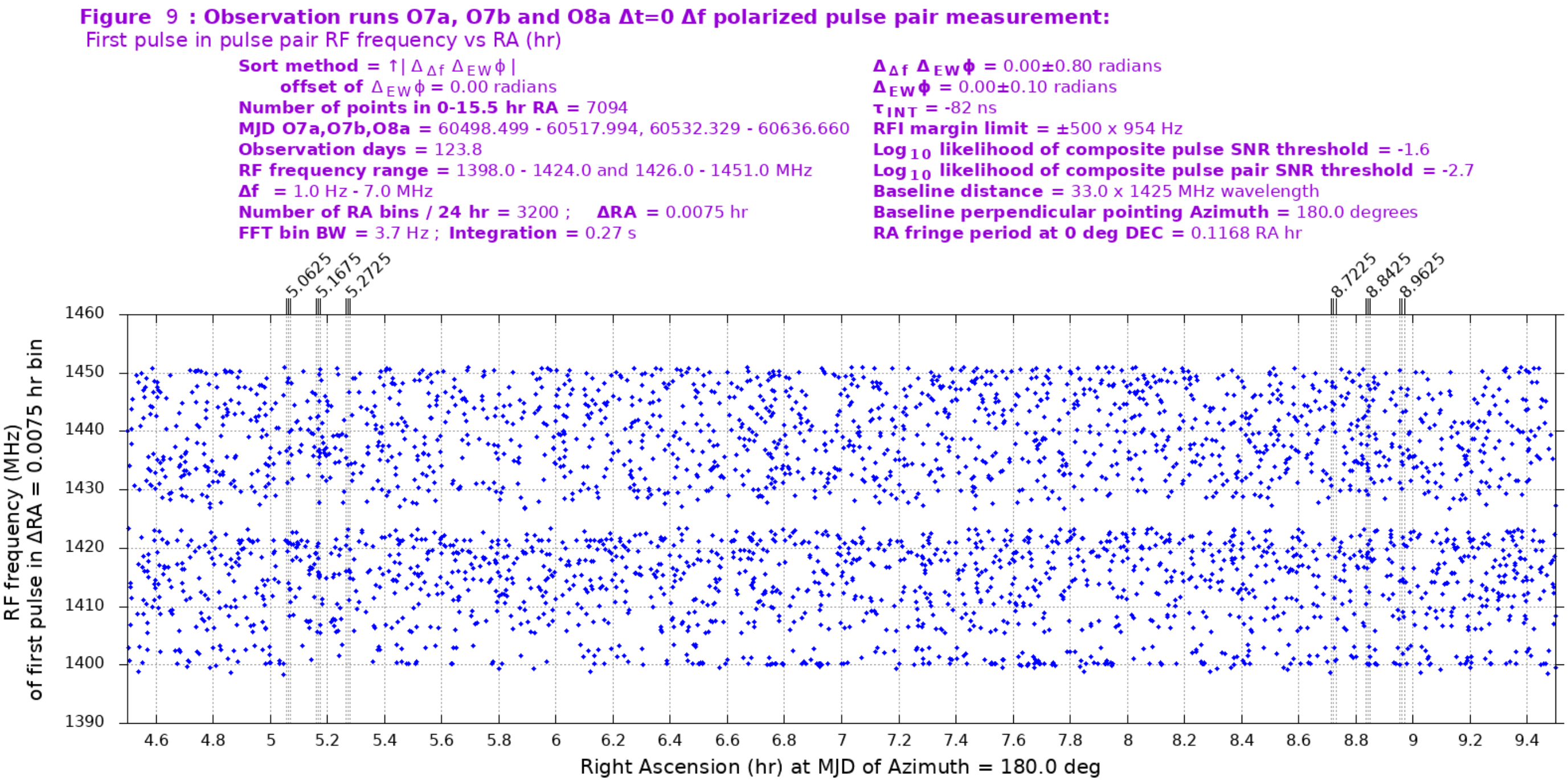

**Figure 9:** Concentrations in RF frequency of pulse pair events are sought to help understand the possible presence of certain types of RFI. Investigation is planned to seek evidence that the 82 ns uncorrected East-West instrument delay, in $\Delta_{EW}\phi$ measurements, is observable, as it might indicate a concentration of pulses at RF frequency offsets of 12.2 MHz or multiples of 12.2 MHz. Phase noise contributions affect this possibility, because the sawtooth expected signal phase shift versus frequency is modified due to the noise of In-phase & Quadrature (IQ) vectors not aligned in phase with IQ signal vectors, for example. The 1424 to 1426 MHz frequency range is filtered out, as low frequency RFI is sometimes present near the center of the IQ baseband. Other regions of sparse data points may be due to the RFI excision algorithm removal of persistent RFI during many days.

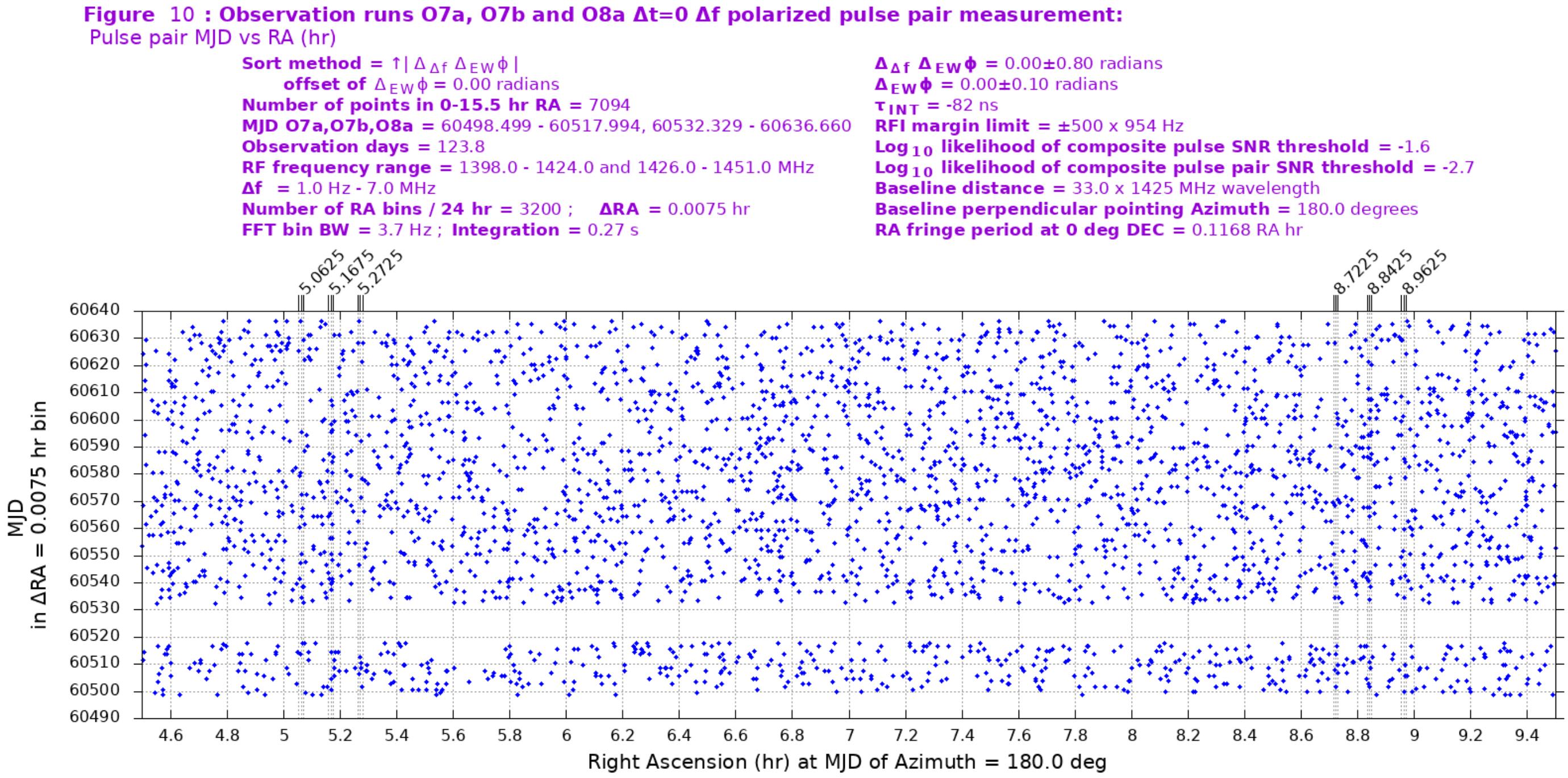

**Figure 10:** Examination of the MJD at the time of a pulse pair event helps in determining if pulse pair events might be explained by intermittent RFI, or equipment issues. The outage in recorded data from MJD 60518.0 to 60532.3 is due to a time period when the receiver system was powered off. Concentration of MJDs in pulse pair events is a topic of further work.

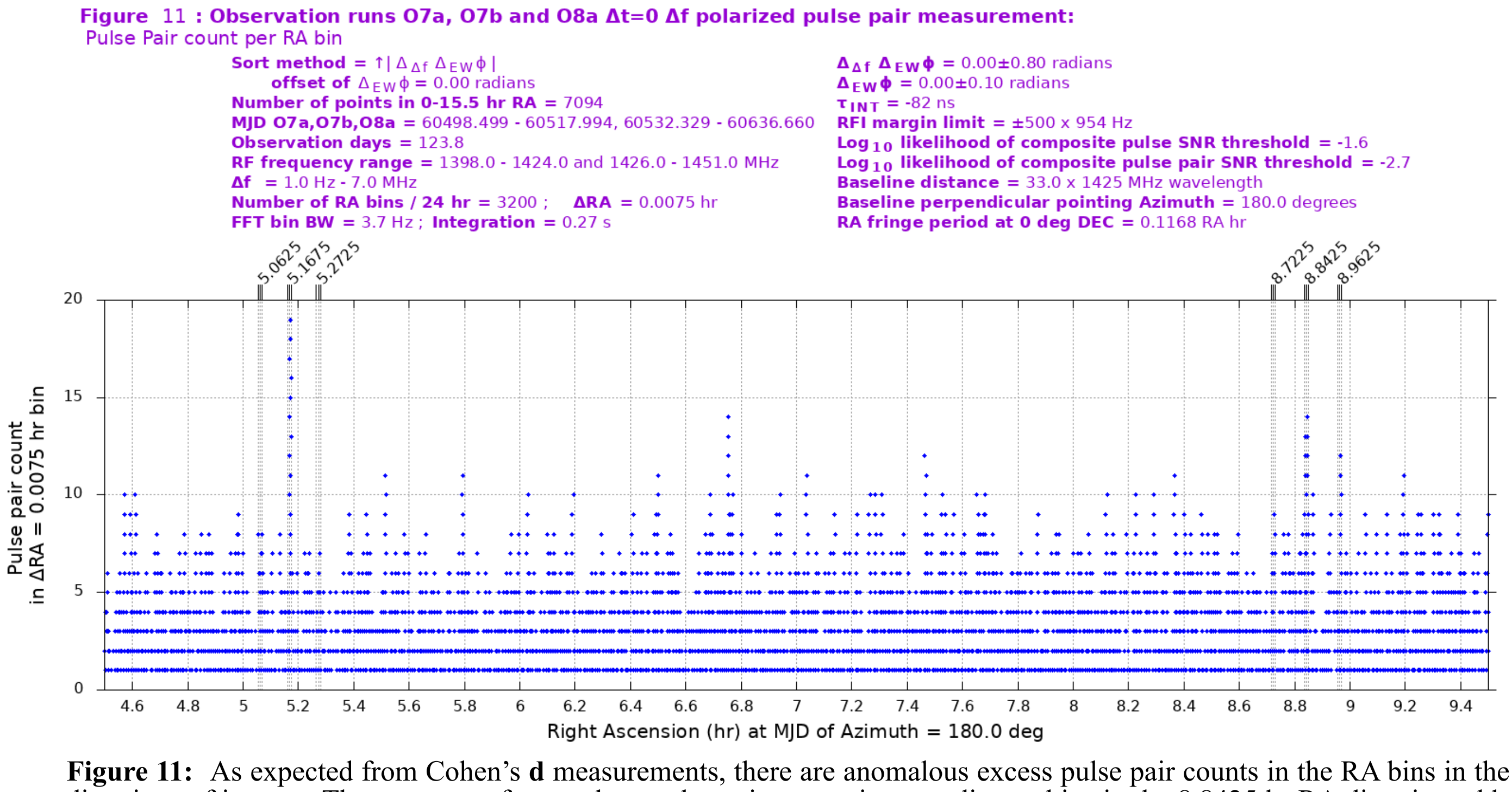

**Figure 11:** As expected from Cohen's **d** measurements, there are anomalous excess pulse pair counts in the RA bins in the two directions of interest. The presence of anomalous pulse pair counts in two adjacent bins in the 8.8425 hr RA direction adds statistical power support to a hypothesis of celestial origin. RA bins correspond to time durations of 27 s per RA bin. Given that only one other RA bin has a count greater than 13, excluding the 5.1675 and 8.8425 RA bins, while assuming the significance of the 8.8425 RA bin, due to its high valued pulse pair count, the likelihood of observing either the lower or upper adjacent RA bins at greater than 12 counts is estimated at 4 RA bins divided by the 666.7 bins in the 5 hr RA range, and calculates to 0.006 likelihood. Cohen's **d** does not have a direct relationship with pulse pair count, because, for example, a small number of counted pulse pairs in an RA bin may have concentrated low values of $|\Delta_{\Delta f}\Delta_{EW}\phi|$, increasing the Cohen's **d** value. Low $|\Delta_{\Delta f}\Delta_{EW}\phi|$ is associated with a common direction of arrival of pulse pair components, thus giving utility to Cohen's **d** effect size measurement of increasing $|\Delta_{\Delta f}\Delta_{EW}\phi|$ across RA populations.

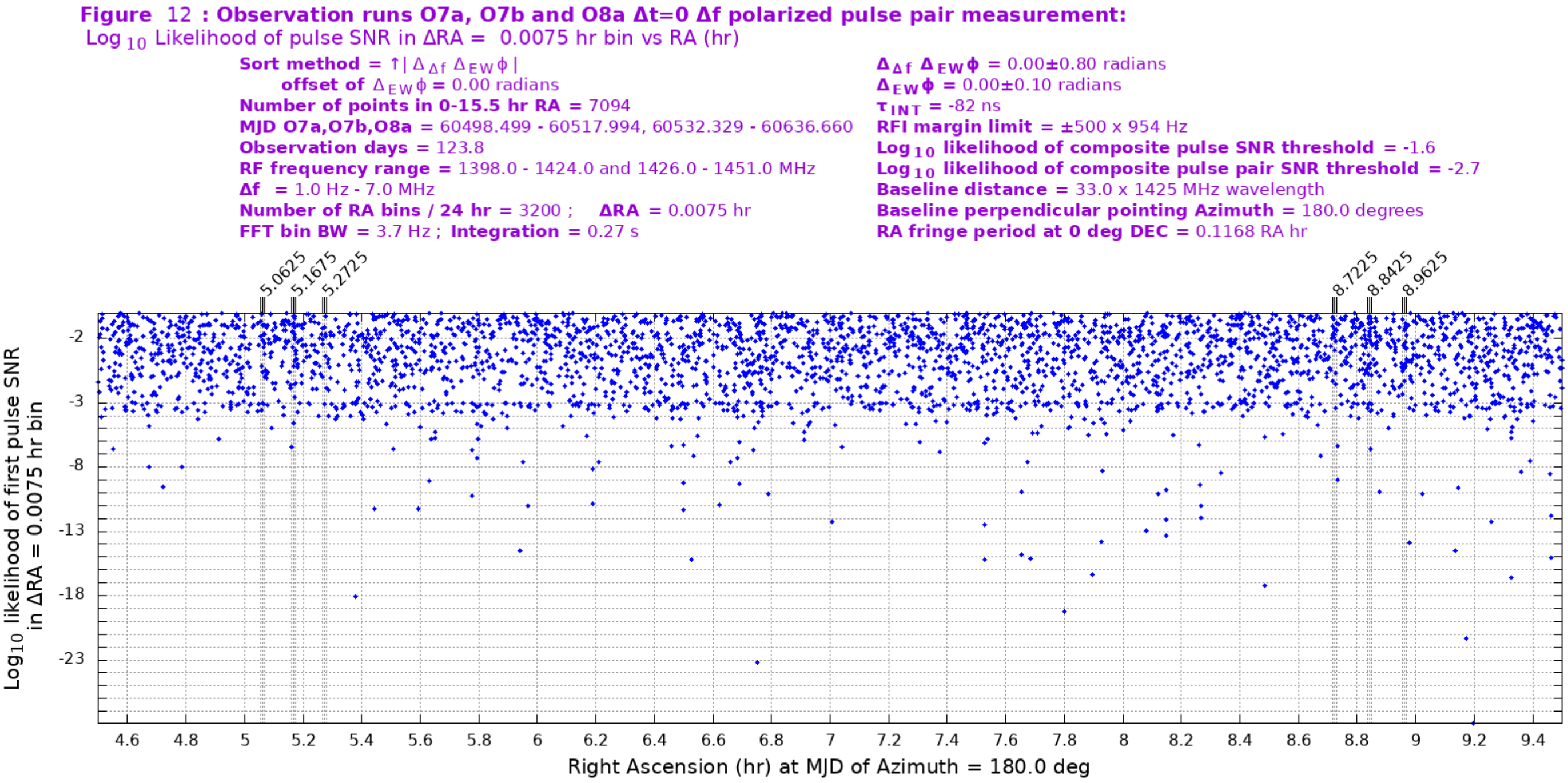

**Figure 12:** $\text{Log}_{10}$ likelihoods of first pulse and pulse pair SNR **[7]** are shown in **Figs. 12-13.** The $\log_{10}$ likelihood values are referenced to a zero value at 8.5 dB pulse SNR. The concentrations near -3 in **Fig. 12** and -5 in **Fig. 13** are due to the $\log_{10}$ likelihood scale change in each plot. $\text{Log}_{10}$ likelihood of SNR is useful to estimate the expected presence of high measured pulse and pulse pair SNR assuming only AWGN is present. Low $\log_{10}$ likelihood values appear sporadically in the comparison RA group due to unknown causes. The sparse presence of low $\log_{10}$ likelihood in the directions of interest implies that the anomalous 3.7 Hz bandwidth pulse pairs observed in the directions of interest are not the result of a powerful or nearby narrow bandwidth emission source. On the other hand, examination of unknown-cause SNR anomalies in the comparison group might help in understanding the Cohen's **d** effect size anomalies in the directions of interest.

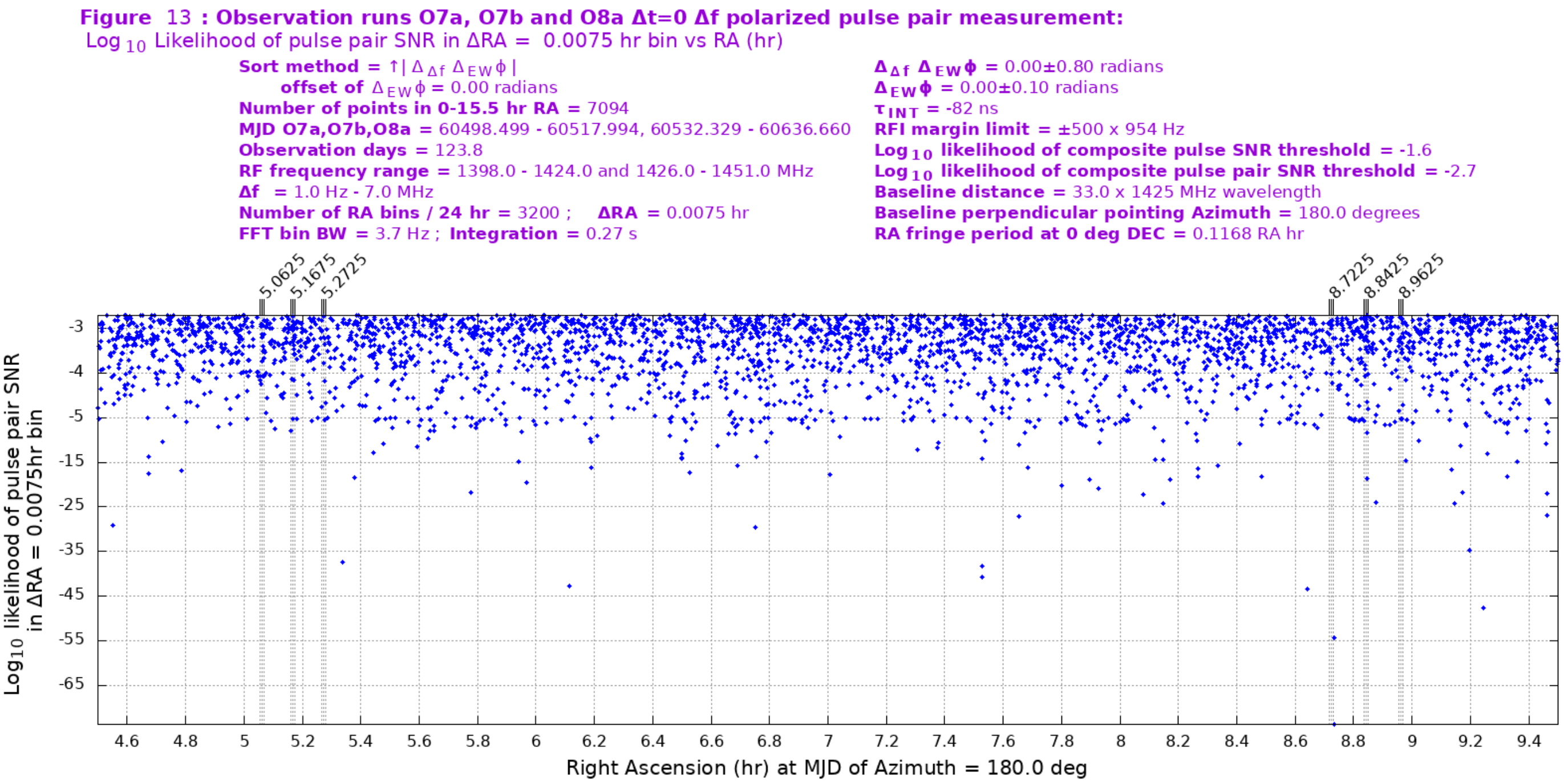

**Figure 13:** $\text{Log}_{10}$ likelihood of the composite SNR of the two pulses in the pulse pairs shows sporadic presence of very unlikely events across RA bins. The presence of outliers at very low likelihood values is not readily expected in AWGN. However, the number of pulses applied to first-level SNR threshold processing during the 123.8 day experiment is very large. There is a possibility that the 954 Hz bandwidth noise power measurement and the 3.7 Hz bandwidth signal power measurement have a mechanism that produces the sporadic low likelihood values. Further work is required to seek explanations.

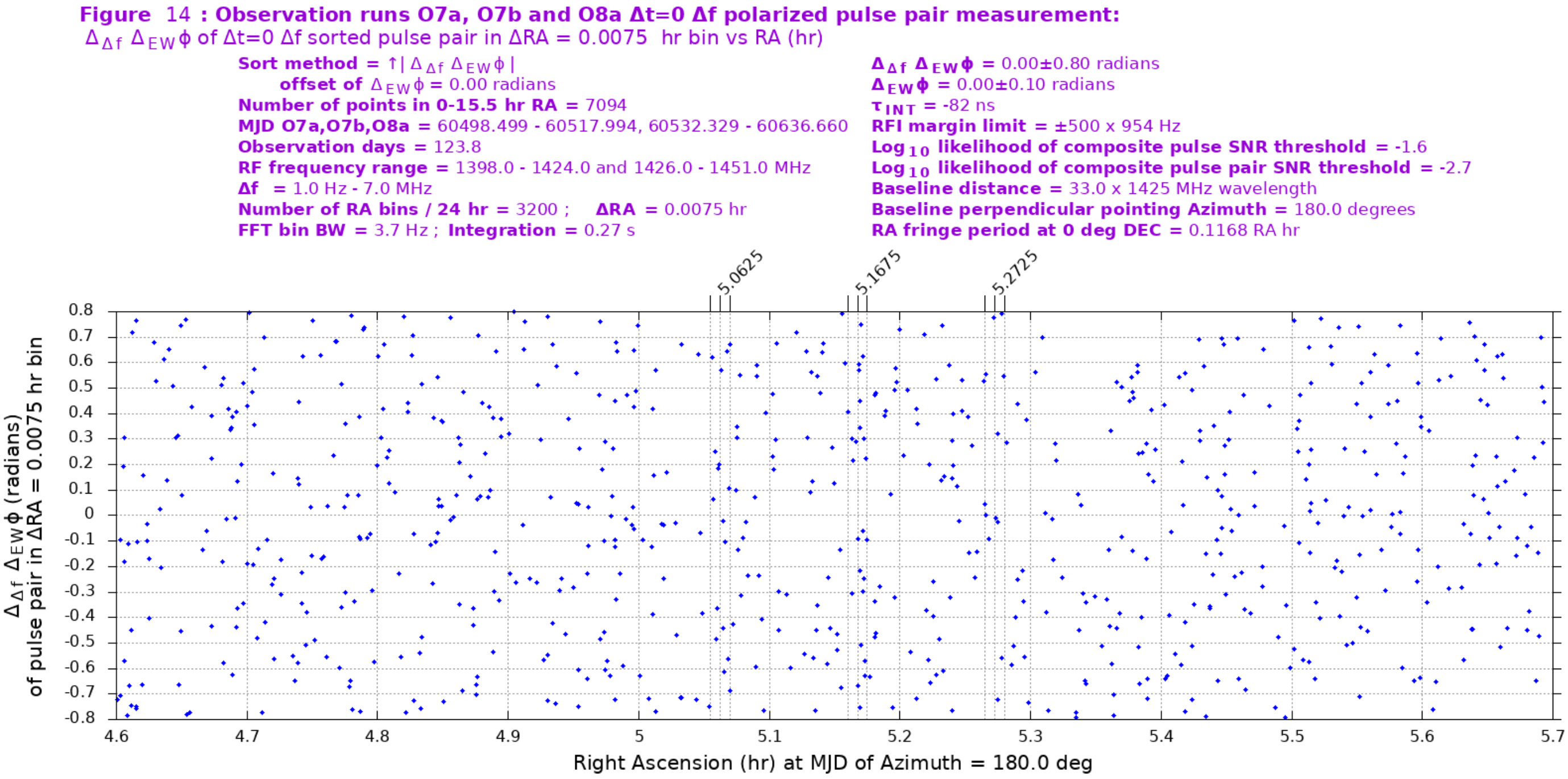

**Figure 14:** The examination of the phase difference of Δf separated pulses in pulse pairs helps to estimate phase noise, if phase noise is less than the second-level processing limits of the $\Delta_{\Delta f}\Delta_{EW}\phi$ measurements. No significant concentration of pulse pair events clearly indicate the possibility of concentrated values, in the prior 5.1-5.4 hr RA direction.

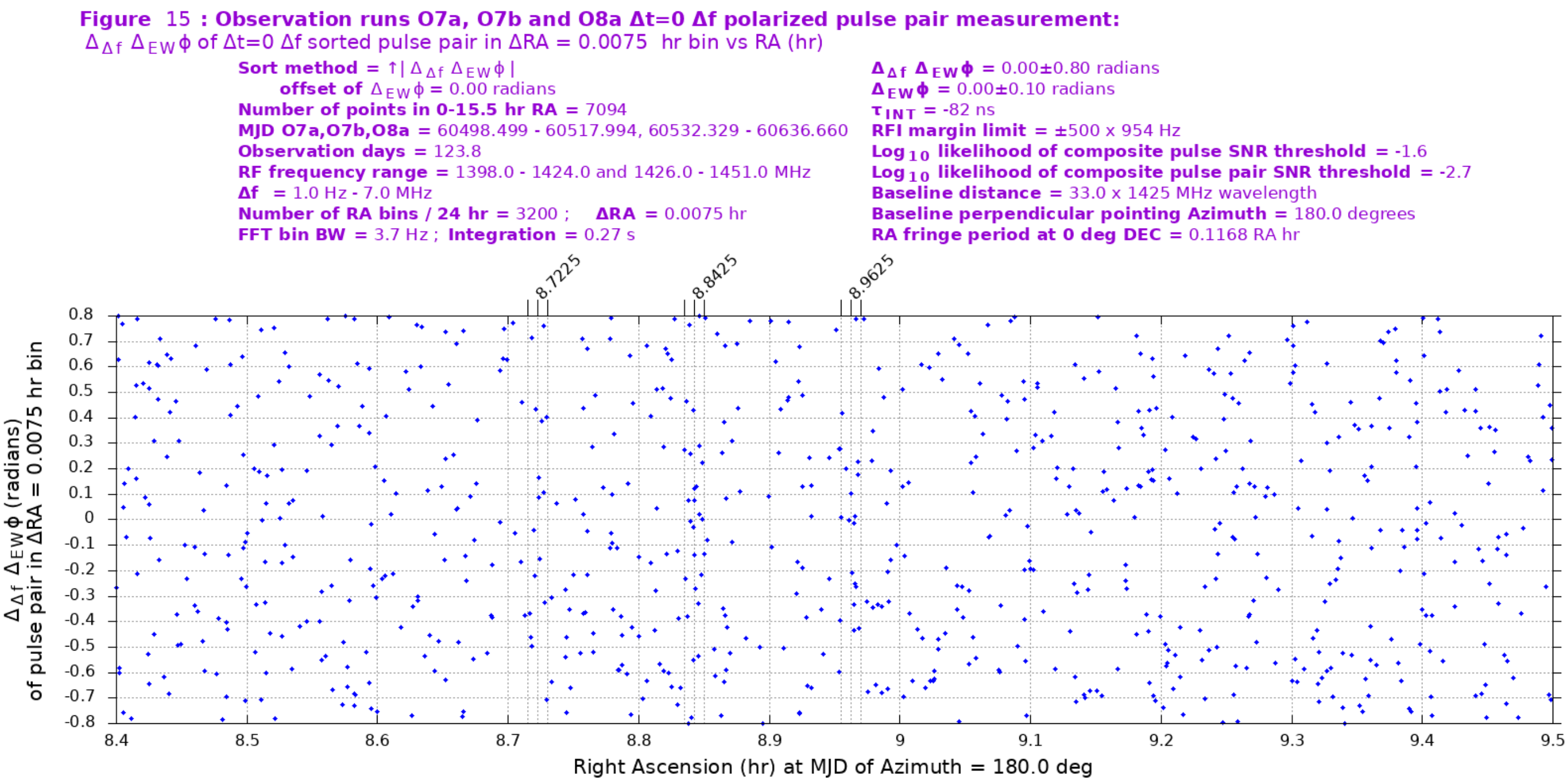

**Figure 15:** There might be a $\Delta_{\Delta f}\Delta_{EW}\phi$ concentration of approximately ±0.5 radian range in the two RA bins centered at 8.8425 hr RA and the upper aliased RA bins. This ±0.5 radian value approximately matches the estimate of expected phase noise at ±0.37 radians calculated in **III. Method of Measurement, Theory of operation**. Pulse pairs that have higher SNR are expected to measure $\Delta_{\Delta f}\Delta_{EW}\phi$ close to zero radians. This property results in a potential RFI filter, because RFI often has wide variation of SNR, and $\Delta_{\Delta f}\Delta_{EW}\phi$. Together, the association of high SNR pulse pairs with high $|\Delta_{\Delta f}\Delta_{EW}\phi|$ values, perhaps at concentrated MJD times, might be used to lead to the identification of RFI, contrasted with celestially-sourced pulse pairs, a topic of ongoing work.

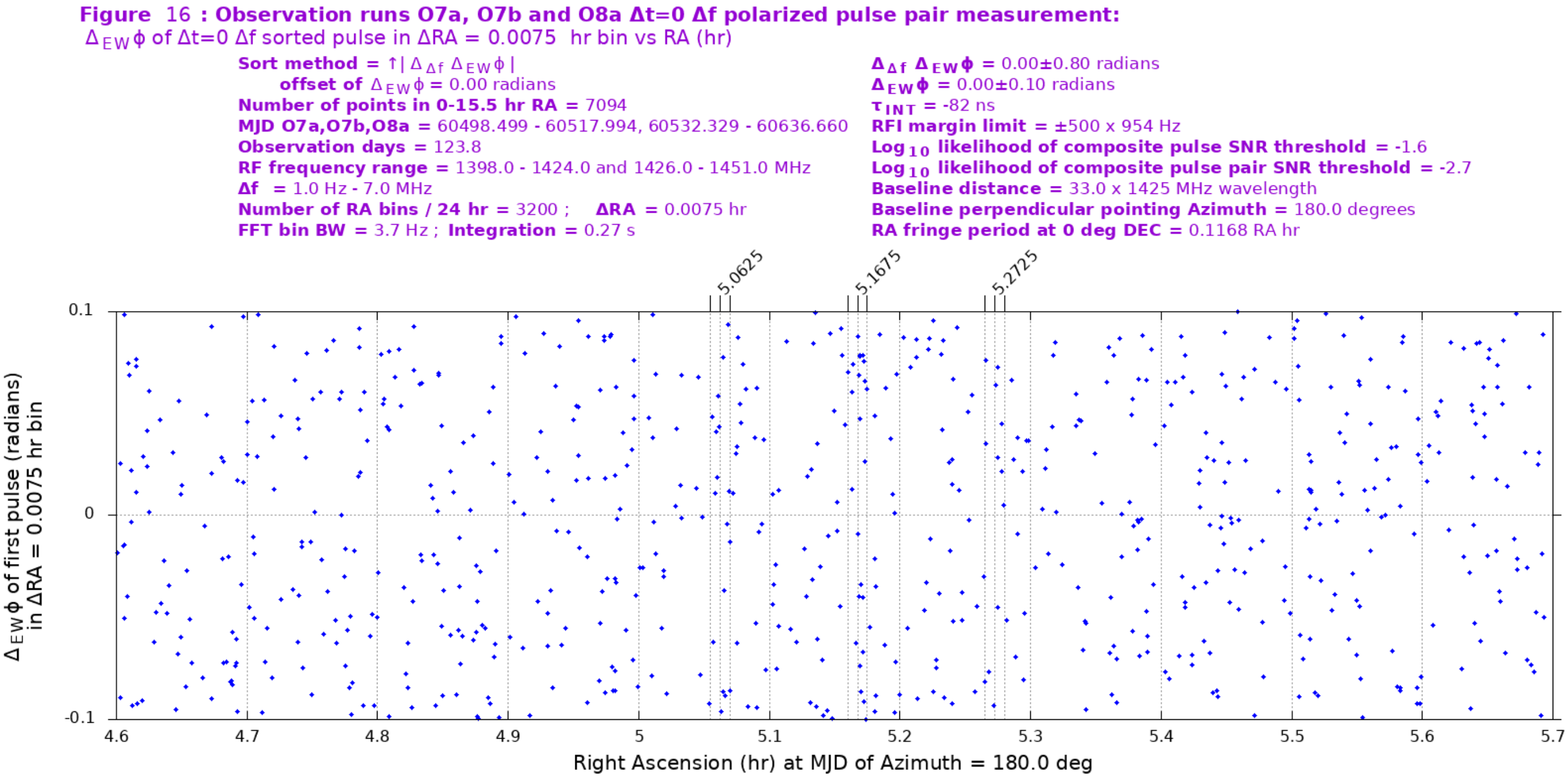

**Figure 16:** A cluster of data points evident near +0.07 radians in the measurement of $\Delta_{EW}\phi$ in the central 5.1675 hr RA bins may be due to a phase detector offset in receiver instrumentation. The RA bin size of ±0.00375 hr RA corresponds to a calculated $\Delta_{EW}\phi$ variation of ±0.2 radian, due to geometric space delay, assuming no phase noise is present, or that phase noise contributions cancel. It seems possible that high SNR received pulses might show concentrations of $\Delta_{EW}\phi$ values in an RA bin associated with the signal's direction of arrival. Large variations in $\Delta_{EW}\phi$ are expected if off-pointing-axis RFI is present.

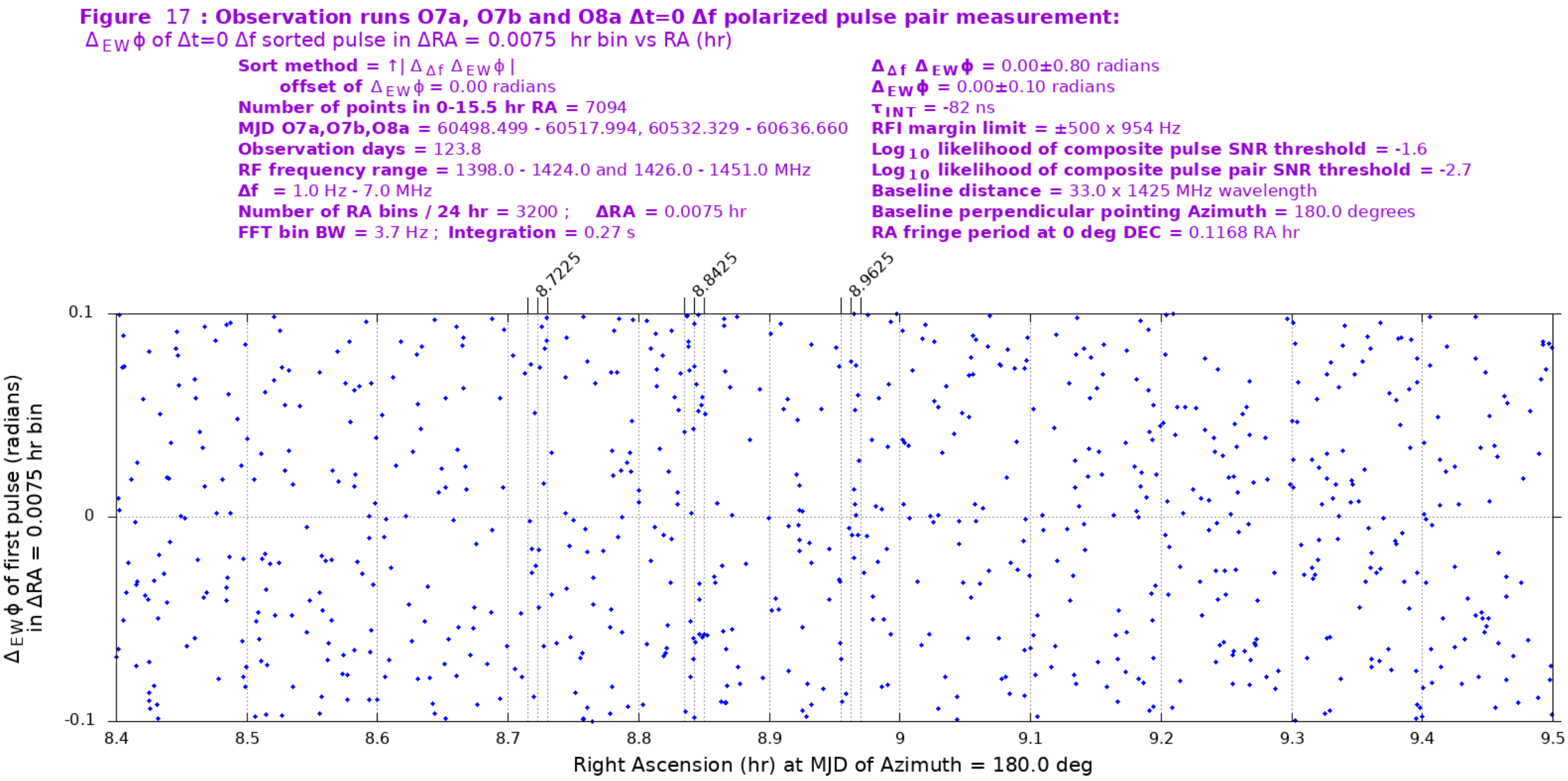

**Figure 17:** The $\Delta_{EW}\phi$ RF phase difference of the lower frequency pulses in the two adjacent hr RA bins centered at 8.8425 hr RA seems to have a bimodal distribution. More data points are required to determine if this is present. The upper 0.120 hr RA aliased RA bin appears to have a significant number of near-zero $\Delta_{EW}\phi$ values, compared to RA bins observed in the comparison group.

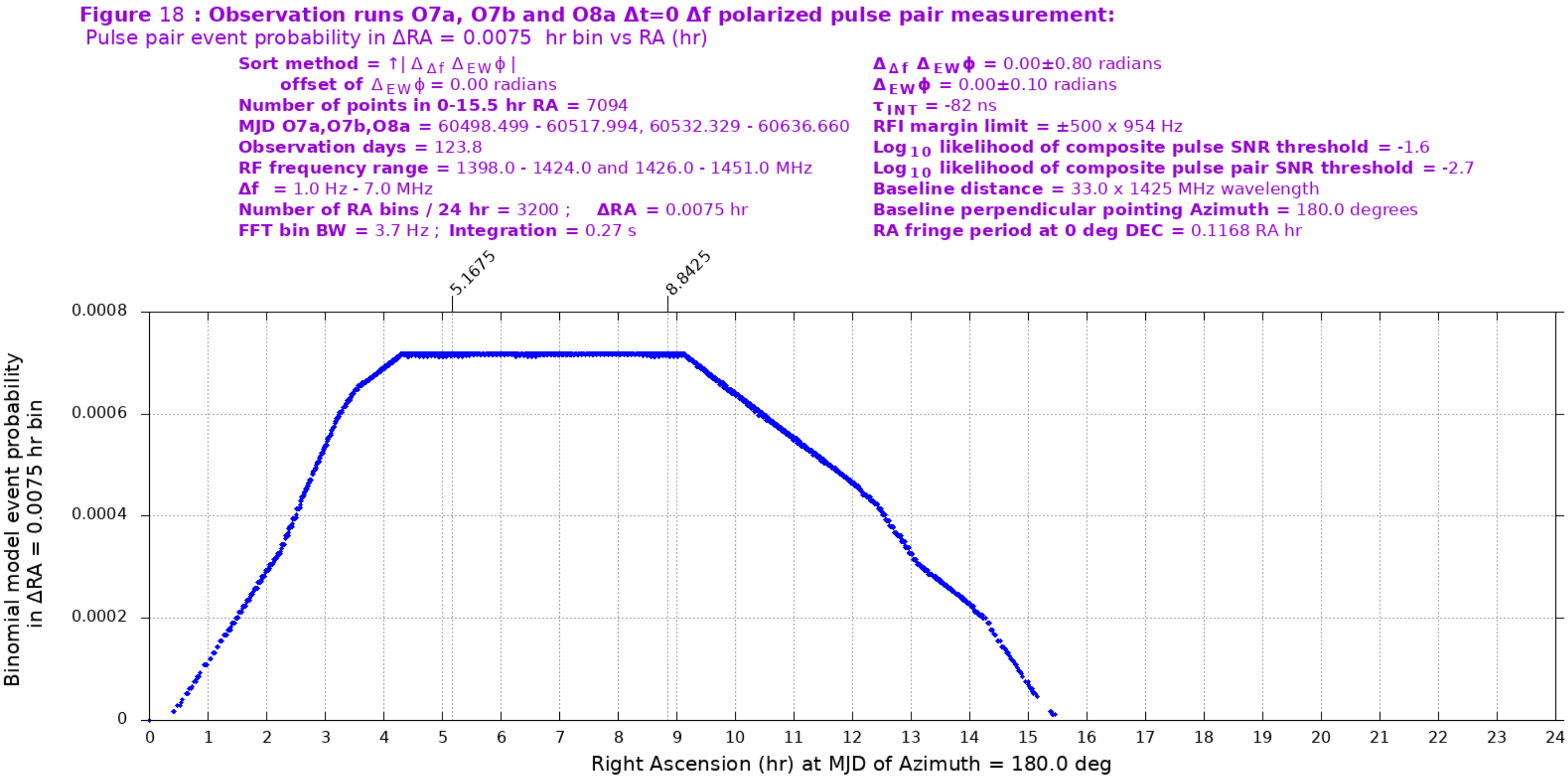

**Figure 18:** The event probability per RA bin depends on the number of first-level processing files used in second-level processing across the RA range. The files each have a four hour MJD duration and are included to flatten the event probability across an examined RA range. Statistical power, using effect size Cohen's **d**, **[9]** is calculated using a binomial modeled process that calculates event probabilities, pulse pair mean counts and standard deviations per RA bin, assuming an AWGN cause. Statistical power may therefore be measured outside of the flattened probability area, albeit with increased Cohen's **d** uncertainty. For example, in **Fig. 20**, the pointing directions of RA 11.978-12.007 hr RA and 11.880-11.887 hr RA, after examination of a second-level processing output data file, show a total of eleven pulse pairs having Cohen's **d** greater than 5 standard deviations, indicating a possibly 0.109 hr RA aliased celestial source direction. Wide RA range increases in pulse pair count effect size are not observed in RA ranges with increased broadband noise. Increasing the number of MJD files is expected to help this investigation.

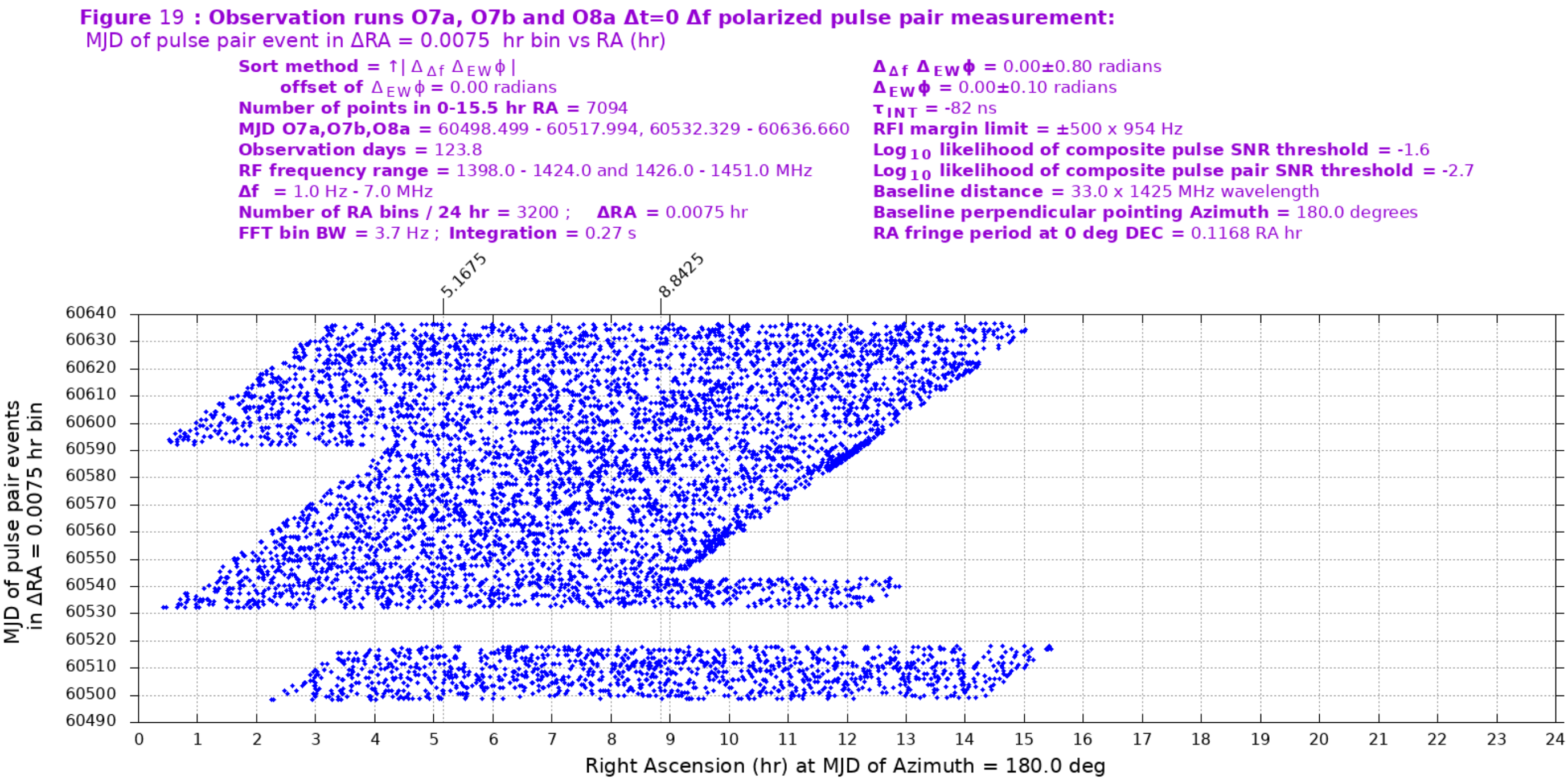

**Figure 19:** All of the MJD days, excluding a powered-off outage period from MJD 60518.0 to 60532.3, are included in second-level processing. The increased concentration of pulse pairs near MJD 60555 at 10 hrs RA, and near MJD 60590 at 12 hrs RA is thought to be due to antenna edge pattern response of the Sun, the latter evident in **Fig. 20**, with MJD 60586 Sun response shown.

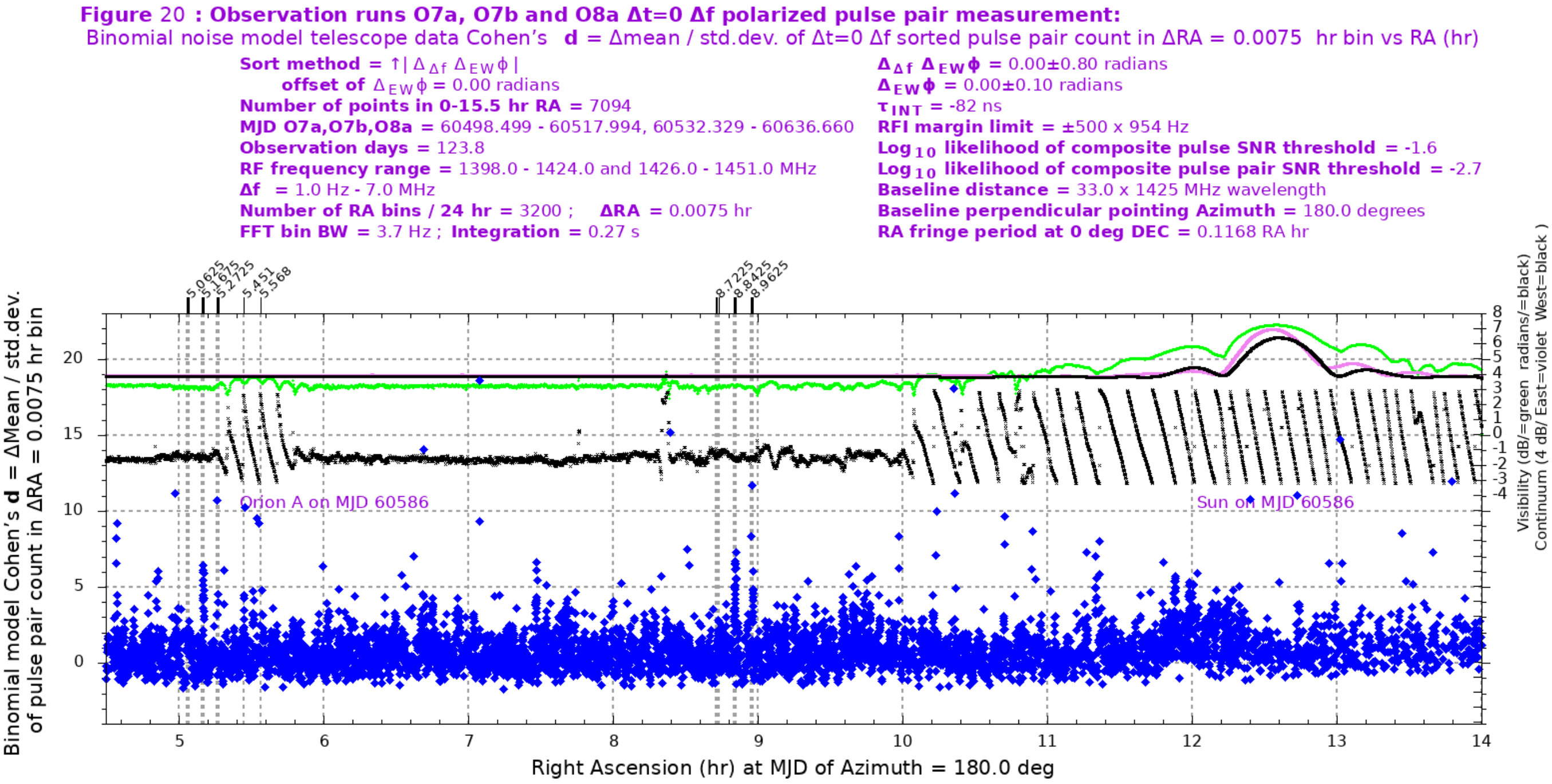

**Figure 20:** The plot in **Fig. 1** is shown in **Fig. 20** with an increase in the RA range, to include the MJD 60586 continuum and 50 MHz bandwidth, 0.27 s integrated, FX correlator response of the Sun transiting the local meridian, having 12.67 hrs RA and -4.3º DEC. A lower resolution scale of continuum power is used. Peak Sun noise measures approximately 10 times system noise. The antenna element beam pointing is estimated at -4.3º±1º DEC, 179.0º±0.5º azimuth, and FWHM 5.3º±1º An increase in level of Cohen's **d** at 12.0±0.3 hr RA is observed at increasing levels of Sun continuum noise measurement, similar to that observed in **Fig. 19**, and thought to result from the increased number of first-level processed pulse pairs at high levels of noise. Several directions, e.g. near 7.47 hr, 11.88 hr and 11.99 hr RA show Cohen's **d** greater than 5 standard deviations, indicating anomalous pulse pair presence in close-spaced RA bins, in addition to the anomalies in the two directions of interest.

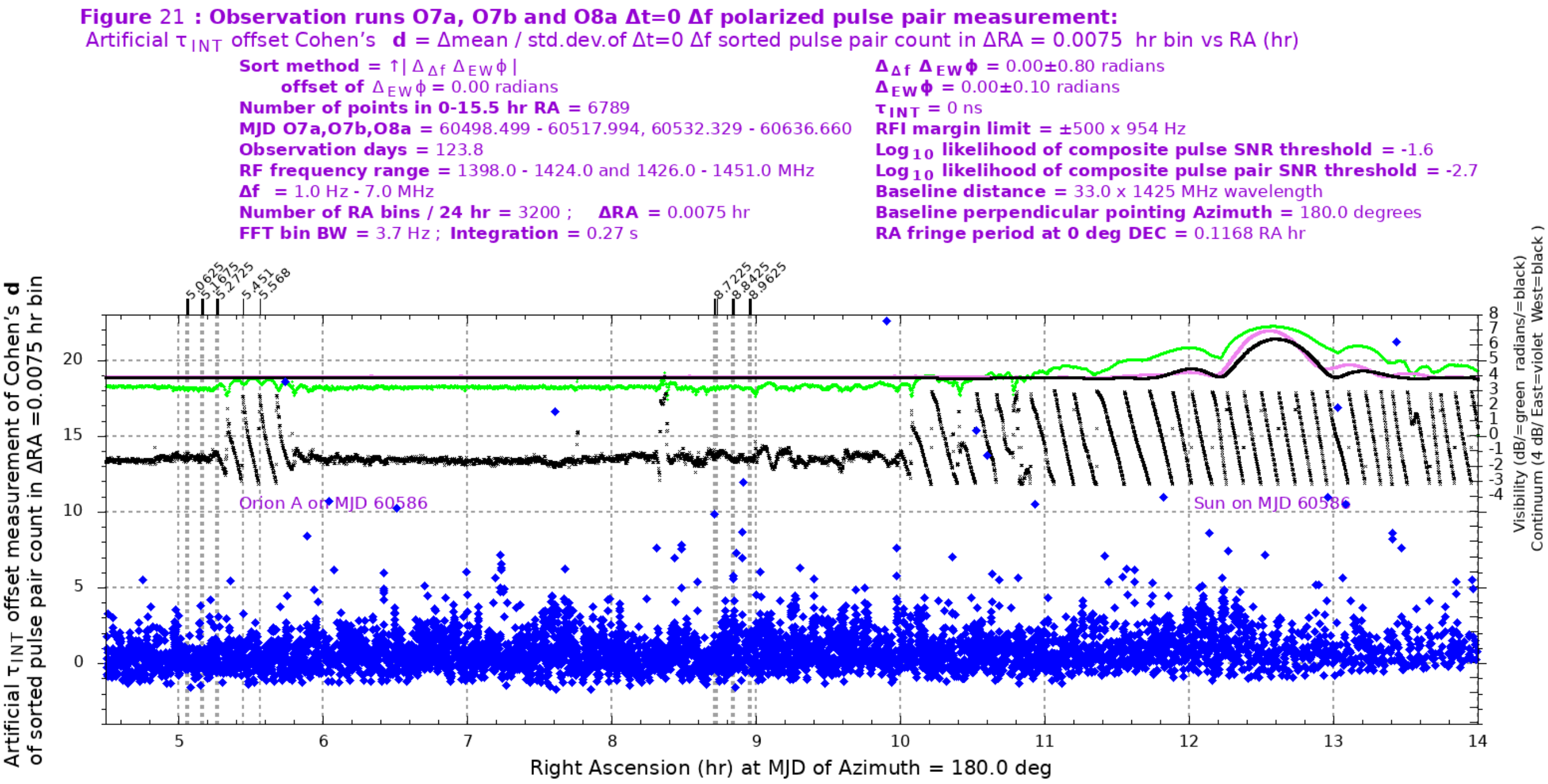

**Figure 21:** One of the potential causes of high values of pulse pair count in RA bins is instrument design issues. The $\tau_{INT}$ measurement at -82 ns compensates the difference in antenna element to phase detector delay, between antenna elements. One way to investigate potential instrument issues is to modify an instrument correction parameter and seek similar anomalous results. Artificially setting the $\tau_{INT}$ value to 0 ns, in second-level processing, prevents Δf-induced delay-corrected pulse pair differential phase from having low values of $\Delta_{\Delta f}\Delta_{EW}\boldsymbol{\phi}$. Cohen's **d** in an RA bin has high sensitivity to low values of $|\Delta_{\Delta f}\Delta_{EW}\boldsymbol{\phi}|$, i.e. those at the top of the sorted heap of candidate pulse pairs. Random pulse pairs at the top of the heap cause sporadic high values of Cohen's **d** without apparent concentrations of pulse pairs in the same or adjacent RA bins. Excluding sporadic outliers, the data indicates that persistent-RA high values of Cohen's **d** at greater than 5 standard deviations are rare, given this system test. The difference in high Cohen's **d** measurements between **Fig. 21** and **Fig. 20** implies that the phenomenon underlying the Cohen's **d** outliers in **Fig. 20** is related to the Δf frequency spacing between components of pulse pairs, and the differential delay between interferometer elements. Examination of potential instrument-induced false positives is a topic of further work.

## V. Discussion

**Noise models**

Assuming that **Fig. 21** is an example of the Cohen's **d** that might result from a noise process, **Figs. 1-7, 11, 19** and **20** show that a noisy process up to approximately telescope system noise, is not a likely explanation for the high statistical power observed in directions of interest.

**Natural object models**

An implication is evidenced in **Fig. 20**, that high levels of pulse pair counts having low $|\Delta_{\Delta f}\Delta_{EW}\phi|$ are not generally expected to be associated with continuum broadband noise power. This implication is deduced using **Figs. 4-7** to also apply to astronomical sources of pulsed broadband energy, e.g. fast radio bursts and pulsars.

Theory and measurements predict that 3.7 Hz bandwidth pulse counts, given an SNR threshold and exponential power probability distributions, do not increase as wide bandwidth noise power level increases, due to the inherent noise-power-normalization in SNR threshold measurements, up to Sun side-lobe added power levels. Thus, many wide bandwidth radiating natural objects do not present favorable models to use to explain excess narrow bandwidth $\Delta t=0$ $\Delta f$ pulse pair count, when high SNR is used to select candidates.

The maximum Doppler shift of a rotating, radiating spherical object is equal to the 3.7 Hz FFT bin bandwidth, if the object has a circumference of 3.7 wavelengths and a rotation rate of one rotation per second. **[7]** The Doppler shift narrow bandwidth desensitization property is one reason why the search for interstellar communication transmitters have used narrow bandwidth receivers. **[12]**

**RFI model falsification attempts**

It seems reasonable to conclude that it is almost impossible to confidently falsify all possible RFI models, because auxiliary hypotheses describing RFI transmitters that match received signals may be imagined. However, over time, RFI transmitter models that are designed to explain long term celestial coordinate repetition of narrow bandwidth pulse pairs are expected to become more difficult to reasonably imagine. For example, in the synchronized, geographically spaced radio telescope prior work at the Green Bank Observatory in West Virginia and the Deep Space Exploration Society in Colorado, the 5.25 hr RA -8º DEC direction was identified as significant.**[1]** A standalone radio telescope was used in subsequent experiments.**[2][3][4]** Repeated significance of the pointing direction of interest, using an interferometer, **[5][6][7]** and in this work, requires a class of RFI models to be designed that avoids RFI model falsification in three different experimental methodologies. For example, such an RFI model needs to be designed to be transmitting pulse pairs during a 27 s interval, for 123.8 days, while a prior celestial direction of interest crosses 180.0º azimuth at the interferometer location, and not at other times.

## VI. Conclusions

The measurement results of this experiment indicate high statistical power, in two celestial pointing directions, of anomalous high counts of $\Delta t=0$ $\Delta f$ polarized 3.7 Hz bandwidth pulse pairs. Measurements in the prior 5.25 ±0.15 hr RA direction and in the 8.8425 hr RA direction, at -4.3º DEC, indicate the presence of unexplained pulse pair count anomalies, presently unlikely to be reasonably explained by known AWGN sources, or known RFI. The hypothesis describing a celestial source is supported by the high number of RA-concentrated pulse count anomalies observed during the 123.8 day experiment. Natural object and intentional narrow bandwidth celestial transmitter hypotheses are not falsified by the findings of this experiment. Many hypotheses remain to be tested. An independently developed system to conduct replication is required before an extraterrestrial intentional transmitter hypothesis may be considered to be more likely than other hypotheses.

## VII. Further Work

1. Continue observations of the two candidate pointing directions, including examination of their aliasing directions, and associated measurements,
2. second-level process other RA directions,
3. study methods of independent replication,
4. examine other DEC pointing,
5. improve instrument metrology,
6. study communication science reasoning to use narrow bandwidth $\Delta t=0$ $\Delta f$ pulse pair transmissions, relative to alternative energy-efficient communication signal discovery methods,
7. add a 3.7 Hz bandwidth measuring receiver for RFI investigation,
8. add a third interferometer element.

## VIII. Acknowledgements

Workers at many organizations have contributed to this long term project, including the Deep Space Exploration Society, Green Bank Observatory, National Radio Astronomy Observatory, SETI Institute, Society of Amateur Radio Astronomers, Breakthrough Listen, Berkeley SETI Research Center, Penn State Extraterrestrial Intelligence Center, Allen Telescope Array, SETI League, HamSCI and hardware and software suppliers. Family and friends have given wonderful ideas and encouragement.

## IX. References

1. W. J. Crilly Jr, *An interstellar communication method: system design and observations*, arXiv: 2105.03727v2, v1 pp. 5, 10, 13, 18, May 8, 2021
2. W. J. Crilly Jr, *Radio interference reduction in interstellar communications: methods and observations*, arXiv: 2106.10168v1, pp. 1, 10, June 18, 2021
3. W. J. Crilly Jr, *Symbol repetition in interstellar communications: methods and observations*, arXiv:2202.12791v1, p. 12, Feb. 25, 2022
4. W. J. Crilly Jr, *Symbol quantization in interstellar communications: methods and observations*, arXiv:2203.10065v1, p. 9, Mar 18, 2022
5. W. J. Crilly Jr, *Interferometer measurements in interstellar communications: methods and observations*, arXiv:2404.08994v1, pp. 2-3, 11, April 13, 2024
6. W. J. Crilly Jr, *Replication of filtered interferometer measurements in interstellar communication*, arXiv:2407:00447v1, pp. 1-3,5, June 29, 2024
7. W. J. Crilly Jr, *A proposed signal discovery method in interstellar communication,* arXiv:2411.02081v1, pp. 2-3,7, 11, Nov. 4, 2024
8. A. R. Thompson, J. M. Moran, G. W. Swenson, Jr., *Interferometry and Synthesis in Radio Astronomy*, Weinheim, FRG: WILEY-VCH Verlag, pp. 50, 192, 290-298, 2004
9. J. Cohen, *Statistical Power Analysis for the Behavioral Sciences*, New York: Academic Press, pp. 19-27, 1977
10. J. D. Kraus, *Radio Astronomy*, Powell, OH: Cygnus-Quasar Books, 2nd Edition, pp. **8**-14, **8**-28, 1966-1986
11. R.F. Bradley, S.A. Heatherly, B. Malphrus, S. Crilly, *Green Bank Observatory 40-Foot Radio Telescope Operator's Manual*, p. 13 Rev. 6/04/2019
12. B. M. Oliver, J. Billingham, Co-Directors, *Project Cyclops*, Stanford University/NASA/Ames Research Center, pp. 56-59, 62, revised ed. July 1973